\documentclass[acmsmall,review,anonymous]{acmart}

\acmJournal{PACMPL}
\acmVolume{1}
\acmNumber{CONF} \acmArticle{1}
\acmYear{2018}
\acmMonth{1}
\acmDOI{} \startPage{1}

\usepackage{listings}
\usepackage{color}
\usepackage{xcolor}
\usepackage{xspace}
\usepackage{graphicx}
\usepackage{subcaption}
\usepackage{amsmath}
\usepackage{cleveref}

\usepackage{marvosym,etoolbox}
\usepackage{mathtools}
\usepackage{hyperref}
\usepackage{mathpartir}
\usepackage{stmaryrd}
\usepackage{enumitem}
\usepackage{multirow}
\usepackage{wrapfig}
\usepackage{float}
\usepackage{makecell}
\usepackage{adjustbox}
\usepackage{balance}
\usepackage{MnSymbol}
\usepackage{algorithm}
\usepackage{algpseudocode}
\usepackage[scientific-notation=true]{siunitx}

\AtBeginDocument{}

\setcopyright{none}
\copyrightyear{2018}
\acmYear{2018}
\acmDOI{XXXXXXX.XXXXXXX}

\acmConference[Conference acronym 'XX]{Make sure to enter the correct
  conference title from your rights confirmation emai}{June 03--05,
  2018}{Woodstock, NY}
\acmPrice{15.00}
\acmISBN{978-1-4503-XXXX-X/18/06}

\usepackage[]{xcolor}

\usepackage{todonotes}

\newcommand{\rn}[1]{\pgwrapper{RRN}{#1}}  

     \newcommand{\mv}[1]{\pgwrapper{MV}{#1}}    \newcommand{\csk}[1]{\pgwrapper{CSK}{#1}}   

\newcommand{\mk}[1]{\pgwrapper{MK}{#1}}  \newcommand{\vs}[1]{\pgwrapper{VS}{#1} } \newcommand{\ap}[1]{\pgwrapper{AP}{#1} }

\InputIfFileExists{activateeditingmarks}{
}{
    \def\noeditingmarks{}
}

\definecolor{comment-red}{rgb}{0.8,0,0}
\definecolor{dark-green}{rgb}{0.0,0.4,0}
\definecolor{dark-blue}{rgb}{0.0,0.0,0.55}
\definecolor{very-dark-green}{rgb}{0.0,0.3,0}

\ifx\noeditingmarks\undefined
\newcommand{\auditme}[1]{{\color{comment-red}{#1}}}

\newcommand{\new}[1]{{\color{very-dark-green}{#1}}}

\definecolor{mygrey}{rgb}{0.7,0.7,0.7}

   \newcommand{\pgwrapper}[2]{\todo[size=\tiny]{#1: #2}}

\InputIfFileExists{activategreybg}{
       \definecolor{comment-red}{rgb}{0.5,0,0}
       \pagecolor{mygrey}
   }{}

\newcommand{\rnfloat}[1]{\pgwrapper{RRN}{#1}\vspace{-4mm}}  

\else

   \newcommand{\auditme}[1]{#1}

   \newcommand{\pgwrapper}[2]{}

   \newcommand{\new}[1]{#1}

   \newcommand{\rnfloat}[1]{}  \fi

\newcommand{\ie}{{i.e.,}}

\def\scrunchit{}
\ifx\scrunchit\undefined
  
\else

\fi

\newcommand{\secref}[1]{Section~\ref{#1}}
\newcommand{\Secref}[1]{\secref{#1}}
\newcommand{\Figref}[1]{Figure~\ref{#1}\xspace}
\newcommand{\figref}[1]{Figure~\ref{#1}\xspace}

\usepackage{upquote}
\usepackage{listings}
\usepackage[]{xcolor}

\definecolor{MidnightBlue}{rgb}{0.1, 0.1, 0.44}
\definecolor{Bittersweet}{rgb}{1.0, 0.44, 0.37}
\definecolor{Cinnamon}{rgb}{0.82, 0.41, 0.12}
\definecolor{ForestGreen}{rgb}{0.0, 0.27, 0.13}
\definecolor{Gray}{rgb}{0.5, 0.5, 0.5}
\definecolor{RoyalPurple}{rgb}{0.47, 0.32, 0.66}
\definecolor{Maroon}{rgb}{0.5, 0.0, 0.0}
\definecolor{OtterBrown}{rgb}{0.4, 0.26, 0.13}
\definecolor{azure}{rgb}{0.0, 0.5, 1.0}
\definecolor{blush}{rgb}{0.87, 0.36, 0.51}

\definecolor{darkgreen}{rgb}{0,0.5,0}
\definecolor{darkred}{rgb}{0.5,0,0}
\definecolor{darkbrown}{rgb}{0.4, 0.26, 0.13}

\definecolor{hsblue}{HTML}{2d3497}
\definecolor{hsgreen}{HTML}{3f7e32}

\definecolor{emacspink}{HTML}{a241f6}
\definecolor{emacsgreen}{HTML}{3c8625}
\definecolor{emacsbrown}{HTML}{a86a2c}
\definecolor{emacsblue}{HTML}{1b38f5}

\lstdefinestyle{haskell}{language=Haskell,
    upquote=true,
    backgroundcolor=\color{white},
    deletekeywords={case,class,data,default,deriving,do,in,instance,let,of,type,where,IO,ST,STM,if,then,else,tail,List,Int},
    morekeywords={[5]class,data,default,deriving,family,instance,type,where,fun,while,global,return,match,for},   morekeywords={[5]in,let,case,of,do,letpacked,letloc,letregion,start,after,if,then,else,spawn,sync,sizeof,cilk_spawn,cilk_sync,continuation_stolen, size_of,tie,addrOf},
    morekeywords={[4]IORef,IO,ST,STM,STRef,TVar},
    morekeywords={[6]Cons,Nil,Int,List,RNG,Ind,Node,Leaf,Empty},
literate=
{::}{{\textcolor{emacsbrown}{::}}}2
        {=}{{\textcolor{emacsbrown}=}}1
        {Lambda}{{$\Lambda$}}1
        {\\\\}{{\char`\\\char`\\}}1
        {>->}{{>->}}3
        {>>=}{>>=}3
        {->}{{$\rightarrow$}}2
        {>=}{{$\geq$}}2
        {<-}{{$\leftarrow$}}2
        {<=}{{$\leq$}}2
        {=>}{{$\Rightarrow$}}2
        {|}{{\textcolor{emacsbrown}{$\mid$}}}1
        {forall}{{{$\pmb{\forall}$}}}1
        {exists}{{$\exists$}}1
        {...}{{$\cdots$}}3        
}

\lstdefinestyle{inline}{style=haskell,
    basicstyle=\footnotesize\ttfamily,
escapechar={\@},
    mathescape=true,
    literate=
        {\\}{{$\lambda$}}1
        {Lambda}{{$\Lambda$}}1        
        {\\\\}{{\char`\\\char`\\}}1
        {>->}{>->}3
        {>>=}{>>=}3
        {->}{{$\rightarrow$\space}}3    {>=}{{$\geq$}}2
        {<-}{{$\leftarrow$}}2
        {<=}{{$\leq$}}2
        {=>}{{$\Rightarrow$}}2
        {|}{{$\mid$}}1
{forall}{{$\forall$}}1
        {exists}{{$\exists$}}1
        {...}{{$\cdots$}}3
}

\lstnewenvironment{code}
    {\lstset{style=haskell}\csname lst@SetFirstLabel\endcsname}
    {\csname lst@SaveFirstLabel\endcsname}
    {}

\newcommand{\il}[1]{\lstinline[style=inline,mathescape=true];#1;}

\newcommand{\makeatchar}{\lstDeleteShortInline@}

\lstdefinestyle{c}{language=C,
    upquote=true,
    escapechar={\@},
    deletekeywords={typedef,struct,char,return,if,else},
    morekeywords={[5]typedef,struct,return,if,else},
    morekeywords={[6]Location,shadowstack_frame,ShadowstackFrame,
                     GibCursorProd3,GibCursor,GibBool,GibInt,GibPackedTag,GibCursorProd3_struct,
                     uint8_t, int64_t, bool, char,uint32_t
                     },
}

\lstnewenvironment{ccode}
    {\lstset{style=c}\csname lst@SetFirstLabel\endcsname}
    {\csname lst@SaveFirstLabel\endcsname}
    {}

\definecolor{canaryblue}{HTML}{2916F5}
\definecolor{darkgreen}{HTML}{006400}
\definecolor{darkpink}{HTML}{CD00CD}
\definecolor{graygreen}{rgb}{0.3,0.5,0.3}
\definecolor{grayblue}{rgb}{0.2,0.2,0.6}
\definecolor{grayred}{rgb}{0.5,0.2,0.2}
\definecolor{crimsonred}{HTML}{990000}
\definecolor{irishgreen}{HTML}{08A04B}
\definecolor{darkviolet}{HTML}{9400D3}

\newcommand{\system}{{\sc Marmoset}\xspace}

\newcommand{\ribbit}{{\sc Ribbit}\xspace}

\newcommand{\systemSolver}{{\sc M$_{\mathit{solver}}$}\xspace}
\newcommand{\systemGreedy}{{\sc M$_{\mathit{greedy}}$}\xspace}

\newcommand{\alignedPre}{{\sc Aligned$_{\mathit{pre}}$}\xspace}
\newcommand{\alignedIn}{{\sc Aligned$_{\mathit{in}}$}\xspace}
\newcommand{\alignedPost}{{\sc Aligned$_{\mathit{post}}$}\xspace}

\newcommand{\misalignedPost}{{\sc Misaligned$_{\mathit{post}}$}\xspace}
\newcommand{\misalignedPre}{{\sc Misaligned$_{\mathit{pre}}$}\xspace}

\newcommand{\systemlang}[0]{\ensuremath{\lambda_{M}}}

\newcommand{\gibbon}{{\sc Gibbon}\xspace}

\newcommand{\relc}{{\sc RELC}\xspace}

\newcommand{\mlton}{{\sc MLton}\xspace}

\newcommand{\ra}[1]{\renewcommand{\arraystretch}{#1}}

\newcommand{\snum}[1]{{\num[round-precision=3,round-mode=figures,scientific-notation=true,output-exponent-marker = e]{#1}}}
\newcommand{\snumfour}[1]{{\num[round-precision=4,round-mode=figures,scientific-notation=true,output-exponent-marker = e]{#1}}}

\newcommand{\fnumthree}[1]{ {\num[round-mode=places,round-precision=3, scientific-notation=false]{#1}}}

\def\cpp{{C\nolinebreak[4]\hspace{-.05em}\raisebox{.4ex}{\footnotesize\bf ++}}}
\def\csharp{C\tikz[x=1em,y=\baselineskip]\draw (0.125,0.15) -- ++(0.15,0.5)(0.325,0.15) -- ++(0.15,0.5)(0.05,0.3) -- ++(0.45,0.0)(0.1,0.5) -- ++(0.45,0.0);}

\newcommand{\gramdef}{\; ::= \;}
\newcommand{\gramor}{\; | \;}
\newcommand{\keywd}[1]{\mathit{#1}}
\newcommand{\gramwd}[1]{\texttt{#1}}

\newcommand{\skeywd}[1]{\mathit{#1}}

\newcommand{\PROG}{\keywd{top}}
\newcommand{\DD}{\keywd{dd}}

\newcommand{\FD}{\keywd{fd}}
\newcommand{\DC}{\keywd{K}}

\newcommand{\fields}{\skeywd{fe}}

\newcommand{\EXPR}{\keywd{e}}

\newcommand{\DATA}{\gramwd{data}}
\newcommand{\TYP}{\keywd{\tau}}

\newcommand{\sTYP}{\skeywd{\tau}}
\newcommand{\var}{\svar}

\newcommand{\svar}{x}
\newcommand{\fvar}{\sfvar}
\newcommand{\sfvar}{f}
\newcommand{\yvar}{y}

\newcommand{\ARROW}{\rightarrow}

\newcommand{\letpack}[3]{\gramwd{let}\;#1=#2\;\gramwd{in}\;#3}

\newcommand{\fappnoloc}[1]{\fvar \; #1}

\newcommand{\pat}{\keywd{pat}}

\newcommand{\ind}{\keywd{i}}

\newcommand{\VAL}{\keywd{v}}

\newcommand{\caseclause}[2]{#1 \;\rightarrow \;#2}
\newcommand{\case}[2]{\gramwd{case}\; #1 \;\gramwd{of}\;#2}

\newcommand{\datacon}[3]{\ensuremath{#1 \;#2 \;#3}}
\newcommand{\dataconnoloc}[2]{\ensuremath{#1 \;#2}}

\newcommand{\tightoverset}[2]{\mathop{#2}\limits^{\vbox to -.5ex{\kern-0.75ex\hbox{$#1$}\vss}}}

\newcommand{\vectorize}[1]{\overline{#1}}

\newcommand{\CFG}[0]{CFG\xspace}

\newcommand{\DFG}[0]{DFG\xspace}

\algnewcommand{\algorithmicswitch}{\textbf{switch}}
\algnewcommand{\algorithmiclet}[1]{\State \textbf{let} #1}
\algnewcommand{\algorithmicletmutable}[1]{\State \textbf{let mutable} #1}
\algnewcommand{\algorithmicmutate}[1]{\State \textbf{mutate} #1}
\algnewcommand{\algorithmicendlet}{\textbf{in}}
\algnewcommand{\algorithmiccase}{\textbf{case}}
\algnewcommand{\algorithmicendcase}{\State \textbf{end}}
\algnewcommand\algorithmicendswitch{\State \textbf{endswitch}}

\algdef{SE}[SWITCH]{Switch}{EndSwitch}[1]{\algorithmicswitch\ #1\ \algorithmicdo}{\algorithmicend\ \algorithmicswitch }
\algdef{SE}[CASE]{Case}{EndCase}[1]{\algorithmiccase\ #1}{\algorithmicend\ \algorithmiccase }

\algnewcommand\Continue{\State \textbf{continue}}
\algblock{Input}{EndInput}
\algnotext{EndInput}
\algblock{Output}{EndOutput}
\algnotext{EndOutput}

\makeatletter
\begin{document}

\title{Optimizing Layout of Recursive Datatypes with Marmoset}
\subtitle{Or, Algorithms ${+}$ Data Layouts ${=}$ Efficient Programs}

\begin{abstract}

While programmers know that the low-level memory representation of data structures can have significant effects 
on performance, compiler support to {\em optimize} the layout of those structures is an under-explored field.
Prior work has optimized the layout of individual, \textit{non-recursive} structures without considering how
collections of those objects in linked or \textit{recursive} data structures are laid out.

\ap{needs rewriting following the new version of Intro}
This work introduces \system, a compiler that optimizes the layouts of algebraic datatypes, with a special focus
on producing highly optimized, {\em packed} data layouts where recursive structures can be traversed with minimal
pointer chasing. \system performs an analysis of how a recursive ADT is used across functions to choose a \textit{global} layout
that promotes simple, strided access for that ADT in memory. It does so by building and solving a constraint system
to minimize an abstract cost model, yielding a predicted efficient layout for the ADT. \system then builds on top
of \gibbon, a prior compiler for packed, mostly-serial representations, to synthesize optimized ADTs. We show experimentally 
that \system is able to choose optimal layouts across a series of microbenchmarks and case studies, outperforming both 
\gibbon's baseline approach, as well as \mlton, a Standard ML compiler that uses traditional
pointer-heavy representations.

\end{abstract}
 \maketitle

\InputIfFileExists{activateeditingmarks}{
    \listoftodos{}
}{
}

\section{Introduction}

Recursive data structures are
readily available in most programming languages. Linked lists, search trees, tries and others
provide efficient and flexible solutions to a wide class of problems---both in low-level languages with direct
memory access (C, \cpp, Rust, Zig) as well as high-level ones (Java, \csharp, Python).
Additionally, in the purely functional (or {\em persistent}~\cite{okasaki}) setting, recursive,
tree-like data structures largely replace array-based ones.
Implementation details of recursive data structures are not necessarily known to
application programmers, 
who can only hope that the library authors and the
compiler
achieve good performance.
Sadly, recursive data structures are a hard optimization target.

High-level languages represent recursive structures with pointers to small
objects allocated sparsely on the heap. An algorithm traversing such a
\emph{boxed} representation spends much time in pointer
chasing, which is a painful operation for
modern hardware architectures. Optimizing compilers for these languages and architectures have
many strengths but optimizing memory representation of user-defined data
structures is not among them. One alternative is resorting to manual memory
management to achieve maximum performance, but it has the obvious drawback of
leaving convenience and safety behind.

\ap{check that we use datatypes v. data types consistently}
A radically different approach is representing recursive datatypes as dense structures
(basically, arrays) \new{with the help of a library or compiler}.
The \gibbon{} compiler tries to improve the performance of recursive data
structures by embracing dense representations by default~\cite{gibbon}. This choice has
practical benefits for programmers: they no longer need to take control of
low-level data representation and allocation to serialize linked structures; and
rather than employing error-prone index arithmetic to access data, they let \gibbon{}
automatically translate idiomatic data structure accesses into operations on
the dense representation.

Dense representations are not a panacea, though. They can suffer a complementary problem due to their inflexibility.
A particular serialization decision for a data structure made by the compiler
can misalign with the behavior of functions accessing that data.
Consider a tree laid out in left-to-right pre-order with a program that accesses that tree right-to-left.
Rather than scanning straightforwardly through the structure, the program would have to jump back and forth
through the buffer to access the necessary data.

\vs{wasn't happy with more-random access as opposed to just random access.}
One way to counter the inflexibility of dense representations is to introduce
some pointers.
For instance, \gibbon{} inserts {\em shortcut pointers} to allow random access to recursive structures~\cite{Local}.
But this defeats the purpose of a dense representation: not only are
accesses no longer nicely strided through memory, but the pointers and pointer
chasing of boxed data are back.
Indeed, when \gibbon{} is presented with a program whose access patterns do not match the chosen data layout,
the generated code can be {\em significantly slower} than a program with favorable access patterns. 

\vs{I think its still faster than a fully pointer-based representation. 
Changing \auditme{than plain, pointer-based representations.} to \auditme{than a program with favorable access patterns}
}

Are we stuck with pointer chasing when processing recursive data structures?
We present \system{} as a counter example. \system{} is our program analysis
and transformation approach that spots misalignments of algorithms and data
layouts and fixes them where possible. Thus, our slogan is:
\begin{quote}
Algorithms + Data Layouts = Efficient Programs
\end{quote}

\system analyzes the data access patterns of a program and synthesizes a data layout that corresponds to that behavior.
It then rewrites the datatype and code to produce more efficient code that
operates on a dense data representation in a way that matches access patterns.
This co-optimization of datatype and code results in improved locality and, in the context of \gibbon,
avoidance of shortcut pointers as much as possible.

We implement \system{} as an extension to \gibbon---a mature
compiler based on dense representations of datatypes.\mv{mature compiler, we wish}\mk{lol}
That way, \system can be either a transparent compiler optimization, or semi-automated tool for exploring different layouts during the programmer's optimization work.
\new{Our approach has general applicability because of the minimal and common nature of the core language:}
the core language of \system (\figref{fig:grammar}) is a simple first-order, monomorphic, strict, purely functional language.
Thanks to this design,\mk{Not a fan of bare ``that''s or ``this''es}
we manage to isolate \system from \gibbon-specific, complicated (backend) mechanics of converting a program
to operate on dense rather than boxed data.

Overall, in this paper:
\begin{itemize}
\item We provide a static analysis capturing the temporal access patterns of a
  function towards a datatype it processes. As a result of the analysis, we
  generate a {\em field access graph} that summarizes these patterns.

\item We define a cost model that, together with the field access graph, enables
  formulating the field-ordering optimization problem as an integer linear program.
  We apply a linear solver to the problem and obtain optimal positions of fields in the
  datatype definition \new{relative to the cost model.}

\item We extend the \gibbon{} compiler to synthesize new datatypes based on the
solution to the optimization problem, and transform the program to use these new,
optimized types, adjusting the code where necessary.
  
\item Using a series of benchmarks, we show that our implementation, \system,
  can provide speedups of 1.14
to 54 times
over the best prior work on dense representations, \gibbon.
\new{\system outperforms \mlton on these same benchmarks by a factor of 1.6 to 38.}

\end{itemize}

\section[Dense Representation]{Dense Representation: The Good, The Bad, and The Pointers}\label{sec:overview}
\label{ssec:overview:dense}
\rn{Lovely section title!}

This section gives a refresher on dense representations of algebraic datatypes (\Secref{sec:overview:overview})
and, using an example, illustrates the performance challenges of picking a
layout for a datatype's dense represention (\Secref{sec:overview:example}).

\subsection{Overview}\label{sec:overview:overview}

\emph{Algebraic datatypes} (ADTs) are a powerful language-based technology. ADTs can
express many complex data structures while, nevertheless, providing a
pleasantly high level of abstraction for application programmers.
The high-level specification of ADTs leaves
space to experiment with low-level implementation strategies. Hence, we use ADTs
and a purely functional setting for our exploration of performance implications of
data layout.

In a conventional implementation of algebraic datatypes, accessing a value of a
given ADT requires dereferencing a pointer to a heap object, then reading the
header word, to get to the payload.
Accessing the desired data may require multiple further pointer dereferences, as objects
may contain pointers to other objects, requiring the unraveling of multiple layers of nesting.
The whole process is often described as \emph{pointer chasing}, a fitting name, 
especially when the work per payload element is low.

In a dense representation of ADTs, as implemented in \gibbon{},
the data constructor stores
one byte for the constructor's {\em tag}, followed immediately by its fields,
in the hope of avoiding pointer chasing.
Wherever possible, the tag value occurs inline in a bytestream that
hosts multiple values.
As a result, values tend to reside compactly in the heap\mk{RAM feels a little low-level for this} using contiguous blocks of
memory. This representation avoids or reduces pointer chasing and admits
efficient linear traversals favored by modern hardware via
prefetching and caching.

\subsection{Running Example}\label{sec:overview:example}

The efficiency of traversals on dense representations of data structures
largely depends on how well access patterns and layout match each other.
Consider a datatype (already monomorphized) describing
a sequence of posts in a blog\footnote{Throughout the paper we use a subset of Haskell syntax, which
corresponds to the input language of the \gibbon compiler.}:
\begin{lstlisting}
    data BlogList = Nil | Blog Content HashTags BlogList
\end{lstlisting}
A non-empty blog value stores a content field (a
string representing the body of the blog post), a list of hash tags
summarizing keywords of the blog post, and a pointer to the rest of
the list.\footnote{In practice, you may want to reuse a standard list type, e.g.:
  \begin{lstlisting}
    type BlogList = [Blog]
    data Blog = Blog Content HashTags
\end{lstlisting}
  but a typical compiler (including
  \gibbon) would specialize the parametric list type with the \lstinline|Blog|
  type to arrive at an equivalent of the definition shown in the main text
  above.}. The datatype has one point of recursion and several variable-length fields in
the definition. To extend on this, \Secref{subsec:eval:trees:rightmost} contains
an example of tree-shaped data (two points of recursion, in particular) with a
fixed-length field. The most general case of multiple points of recursion and
variable-length fields is also handled by \system.

\begin{figure}[]
  \includegraphics[width=0.4\textwidth]{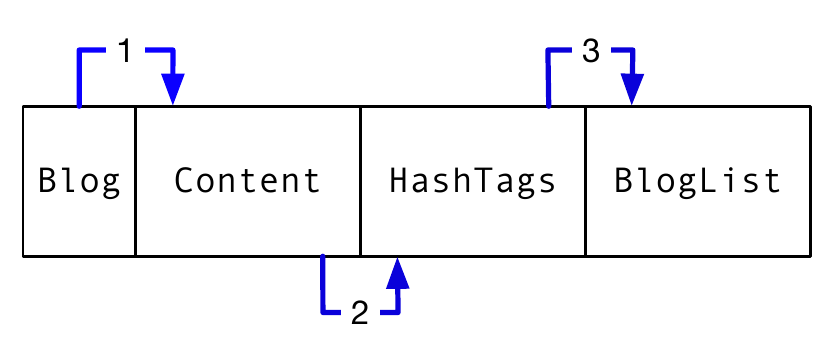}
  \caption{Dense pointer-free layout with ideal access patterns. \new{Numbers represent access order.}}
  \label{fig:traversal-ideal}
\end{figure}

The most favorable traversal for the \lstinline|Blog| datatype
is the same as the order in
which the fields appear in the datatype definition. In this case, \gibbon{} can
assign the dense and pointer-free layout as shown in Figure~\ref{fig:traversal-ideal}.
Solid blue arrows connecting adjacent fields represent unconditional sequential
accesses---i.e., reading a range of bytes in a buffer, and then reading the next
consecutive range.
Such a traversal will reap the benefits of locality.

\begin{figure}[]
    \begin{minipage}[c]{\linewidth}
    \centering
    \subfloat[][Then branch]{
    \includegraphics[width=0.4\textwidth]{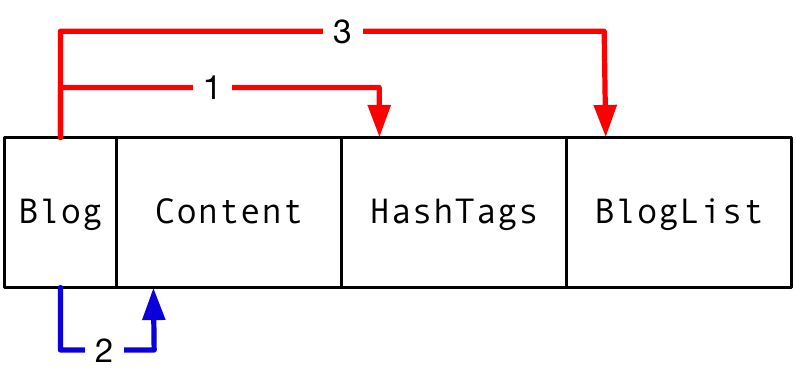}\label{fig:traversal-blog-time-complexity-left}}
    \subfloat[][Else branch]{
    \includegraphics[width=0.4\textwidth]{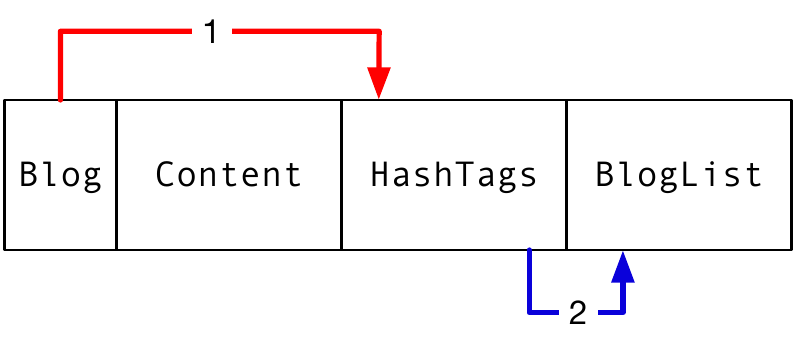}\label{fig:traversal-blog-time-complexity-right}}
    \label{fig:traversal-blog-time-complexity}
    \caption{\new{Out of order accesses (red), incurring costly extra traversals over fields in the middle.}}
  \end{minipage}\hfill
\end{figure}

On the other hand, consider a traversal with less efficient access patterns (\figref{fig:blog-traversal}).
The algorithm scans blog entries for a given keyword in \lstinline|HashTags|.
If the hash tags of a particular blog entry contain the keyword, the
algorithm puts an emphasis on every occurrence of the keyword in the content
field.
In terms of access patterns, if we found a match in the hash tags field,
subsequent accesses to the fields happen in order, as depicted in Figure~\ref{fig:traversal-blog-time-complexity-left}.
Otherwise, the traversal skips over the content field, as depicted in Figure~\ref{fig:traversal-blog-time-complexity-right}.

Here we use red lines to represent accesses that must \emph{skip over} some data
between the current position in the buffer\mk{I don't think we introduce the term ``cursor'' here} and the target data. \new{Data may be constructed recursively
and will not necessarily have a statically-known size, so finding the end of a piece of unneeded data requires
scanning through that data in order to reach the target data.}
This \new{extra traversal (parsing data just to find the end of it)} requires an arbitrarily large amount of work because
the \lstinline|content| field has a variable, dynamically-allocated size.

\begin{figure}[]
  \begin{lstlisting}[]
  emphKeyword :: String -> BlogList -> BlogList
  emphKeyword keyword blogs = case blogs of
      Nil -> Nil
      Blog content hashTags blogs' ->
          case search keyword hashTags of
            True -> let content' = emphContent content keyword
                        blogs''  = emphKeyword keyword blogs'
                      in Blog content' hashTags blogs''
            False -> let blogs'' = emphKeyword keyword blogs'
                       in Blog content hashTags blogs''
   \end{lstlisting}
  \caption{Blog traversal motivating example}
  \label{fig:blog-traversal}
\end{figure}

\new{Extra traversals that perform useful no work}, like skipping over the \lstinline|content| field above,
can degrade the asymptotic efficiency of programs.
One way to avoid such traversals is to use pointers.
For instance, when \gibbon detects that it has to skip over intervening data, it
changes the definition of the constructor by inserting \emph{shortcut pointers},
which provide an exact memory address to skip to in constant time.
For our example program, \gibbon{} introduces one shortcut
pointer for the \lstinline|HashTags| field, and another one for the tail of the list.

  \begin{figure}[]
    \begin{minipage}[c]{\linewidth}
    \centering
    \subfloat[][Then branch]{
    \includegraphics[width=0.4\textwidth]{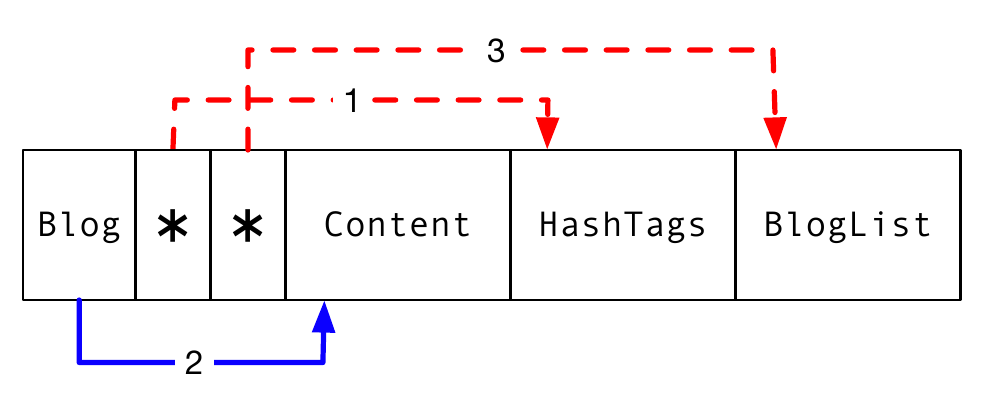}\label{fig:traversal-blog-pointers-left}}
    \subfloat[][Else branch]{
    \includegraphics[width=0.4\textwidth]{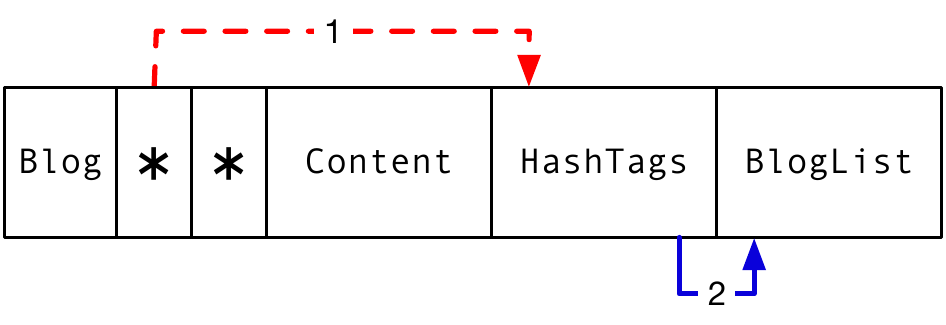}\label{fig:traversal-blog-pointers-right}}
    \caption{Representation with pointers to allow random access.}
    \label{fig:traversal-blog-pointers}
  \end{minipage}\hfill
  \end{figure}

The pointers provide direct accesses to the respective fields when needed and
restore the constant-time asymptotic complexity for certain operations.
\new{This results in the access patterns shown in Figure~\ref{fig:traversal-blog-pointers}.}
Red dashed lines represent pointer-based constant-time field accesses.
Otherwise, the access patterns are similar to what we had before introducing the
pointers.

The pointer-based approach in our example has two weaknesses. First, this
approach is susceptible to the usual problems with pointer chasing. Second, just
like with the initial solution, we access fields in an order that does not match
the layout: the hash tags field is always accessed first but lives next to the
content field.

\begin{figure}[]
  \begin{minipage}[c]{\linewidth}
  \centering
  \subfloat[][Then branch]{
  \includegraphics[width=0.4\textwidth]{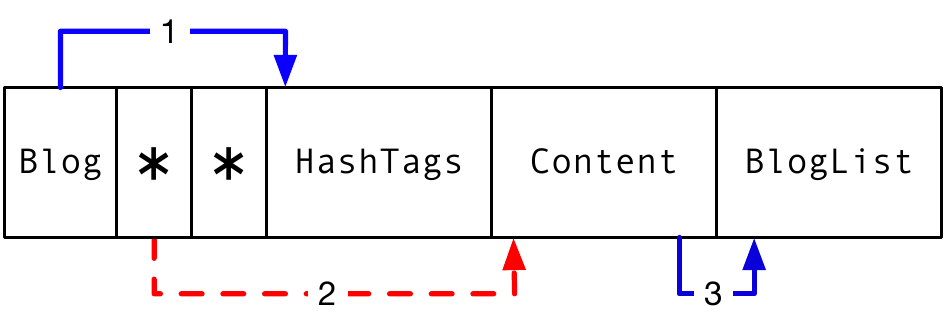}}
  \subfloat[][Else branch]{
  \includegraphics[width=0.4\textwidth]{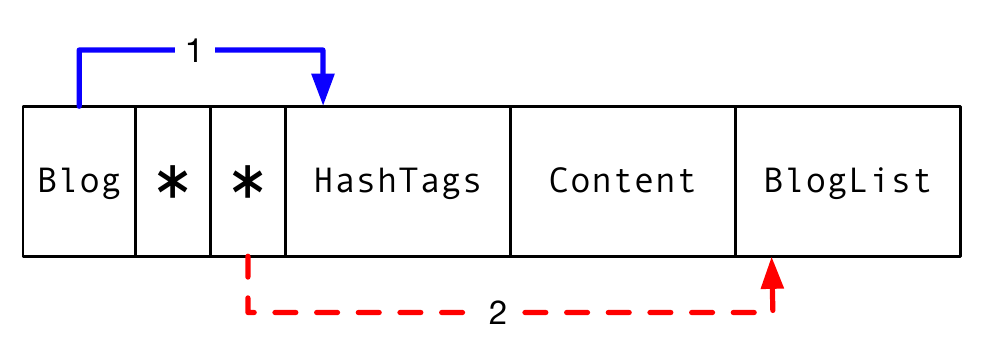}}
  \caption{Optimized layout with favorable access patterns.}
  \label{fig:traversal-blog-optimized}
\end{minipage}\hfill
\end{figure}

\system, described in the following section, automates the process of finding
the weaknesses of the pointer-based approach and improving the layout and code accordingly.
For instance, performance in our example can be improved by swapping the
ordering between the \lstinline|Content| and \lstinline|HashTags| fields.
Given this reordering, the hash tags are available directly at the start of the
value, which lines up with the algorithm better, as the algorithm always
accesses this field first.
Additionally, our program's \lstinline|True| case (the keyword gets a hit within the
hash tags) is more efficient because after traversing the content to highlight
the keyword it stops at the next blog entry ready for the algorithm to make
the recursive call.
This improved data layout results in the more-streamlined access patterns shown in Figure~\ref{fig:traversal-blog-optimized}.

\rn{It would be nice to label these with ``THEN/ELSE'' or ``HIT/MISS'' or something}

 \section{Design}
\label{sec:design}

\system infers efficient layouts for dense representations of recursive
datatypes. \system's key idea is that the best data layout should match the way
a program accesses these data.
\secref{sec:overview} shows how this idea reduces to finding an
ordering of fields in data constructors. The ordering must align with the
order a function accesses those fields, in which case the optimization improves
performance of the function.

To find a better layout for a datatype in the single-function case, \system first analyzes possible executions of the function and their potential for field accesses. In particular, \system takes into account (a) the various paths through a function, each of which may access fields in a different order, and (b) dependencies between operations in the function, as in the absence of dependencies, the function can be rewritten to access fields in the original order, and that order will work best. \system thus constructs a control flow graph (\secref{design:cfg}) and collects dataflow information (\secref{design:dfa}) to build a {\em field access graph}, a representation of the various possible orders in which a function might access fields (\secref{subsec:attributes} and \secref{design:fieldgraph}).

Once data accesses in a function are summarized in the field access graph, \system proceeds with synthesizing a data layout. \system incorporates knowledge about the benefits of sequential, strided access and the drawbacks of pointer chasing and backtracking to define an abstract cost model. The cost model allows to formulate an integer linear program whose optimal solution corresponds to a layout that minimizes the cost according to that model (\secref{design:genconstraints}).

The remainder of this section walks through this design in detail, and discusses how to extend the system to handle multiple functions that use a datatype (\secref{design:global}).

\subsection{\system's Language}
\system{} operates on the language \systemlang{} given in \figref{fig:grammar}.
\systemlang{} is a first-order, monomorphic, call-by-value functional language
with algebraic datatypes and pattern matching.
Programs consist of a series of datatype definitions, function definitions, and a main expression.
\systemlang{}'s expressions use A-normal form~\cite{ANF}.
The notation $\vectorize{x}$ denotes a vector $ [x_1, \ldots, x_n] $ and $\vectorize{x_{\ind}}$
the item at position $\ind$.
\systemlang{} is an intermediate representation (IR) used towards the front end in the Gibbon compiler.
The monomorphizer and specializer lower a program written in a polymorphic, higher-order subset of Haskell\footnote{With strict evaluation semantics using \il{-XStrict}.} to \systemlang{},
and then location inference is used to convert it to the {\em location calculus} (LoCal) code next~\cite{Local}.
It is easier to update the layout of types in \systemlang{} compared to LoCal,
as in \systemlang{} the layout is implicitly determined by the ordering of fields,
whereas the later LoCal IR makes the layout explicit using locations and regions (essentially, buffers and pointer arithmetic).

\begin{figure}[]
	\small

\begin{displaymath}
  \begin{aligned}
  &\DC \in \; \textup{Data Constructors},\:\: \TYP \in \; \textup{Type Constructors},
  \:\: \var, \yvar,\fvar \in \; \textup{Variables}\end{aligned}
\end{displaymath}
\begin{displaymath}
  \begin{aligned}
    \textup{Top-Level Programs} && \PROG && \gramdef & \vectorize{\DD} \;; \vectorize{\FD} \;; \EXPR \\
    \textup{Field of a Data Constructor} && \fields && \gramdef & \sTYP  \\
    \textup{Datatype Declarations} && \DD && \gramdef & \DATA\;\TYP = \vectorize{\DC \; \vectorize{\fields}\;} \\
    \textup{Function Declarations} && \FD && \gramdef & \fvar : \vectorize{\TYP} \ARROW \TYP ; \fappnoloc{\vectorize{\var}} = \EXPR \\
\textup{Values} && \VAL && \gramdef & \var \\ \textup{Expressions} && \EXPR && \gramdef & \VAL\\[-5pt]
&& && \gramor & \letpack{\var:\TYP}{\fappnoloc{\vectorize{\VAL}}}{\EXPR} \\
    && && \gramor & \letpack{\var:\TYP}{\dataconnoloc{\DC}{\vectorize{\VAL}}}{\EXPR} \\
&& && \gramor & \case{\VAL}{\vectorize{\pat}} \\
    \textup{Pattern} && \pat && \gramdef &
    \caseclause{\datacon{\DC}{}{(\vectorize{\var : \TYP})}}{\EXPR} \\
\end{aligned}
\end{displaymath}
 	\normalsize
        \caption{\new{Simplified grammar for a first-order, monomorphic functional language
            that \system operates on. \auditme{The full implementation includes more primitive datatypes, and built in primitives (such as a copy primitive).}}}
\label{fig:grammar}  
\end{figure}

\subsection{Control Flow Analysis}
\label{design:cfg}

\mv{we don't really need an aside about how other compilers for imperative languages format their
CFGs differently}

\begin{algorithm}
    \caption{Control Flow Graph Psuedocode}\label{alg:cfg}
    \begin{algorithmic}[1]
		\small
		\Input 
			\State exp: An expression in subset of \systemlang
			\State weight: The likelyhood of exp executing (\ie{} exp's inbound edge)
\EndInput
		\Output
			\State A tuple of list of cfg nodes and the node id.
		\EndOutput
        \Function{ControlFlowGraph}{exp, weight}
            \algorithmiclet{nodeId = genFreshId()}
            \Switch{$exp$}
                \Case{LetE (v, ty, rhs) bod} \label{alg:cfg-let} 
					\algorithmiclet{(nodes, succId) = \Call{ControlFlowGraph}{bod, weight}}
					\algorithmiclet{newNode = (nodeId, (LetRHS (v, ty, rhs), weight), [succId])} \label{alg:cfg-likelihood}
					\State \Return (nodes ++ newNode, nodeId)
                \EndCase
\Case{CaseE scrt cases} \label{alg:cfg-case}
					\algorithmiclet{(nodes, successors) = \Call{CfgCase}{weight/length(cases), cases}}
					\algorithmiclet{newNode = (nodeId, (scrt, weight), successors)}
					\State \Return (nodes ++ [newNode], nodeId)
				\EndCase
                \Case{VarE v}
                                        \algorithmiclet{newNode = (nodeId, (v, weight), [])}
					\State \Return ([newNode], nodeId)
                \EndCase
            \EndSwitch
        \EndFunction
    \end{algorithmic}
\end{algorithm}

We construct a control flow graph with sub-expressions, and let-bound RHS's (right hand sides) of \systemlang{} as the nodes.
Algorithm~\ref{alg:cfg} shows the psuedocode for generating the control flow graph.
\new{Because the syntax is flattened into A-normal form, there is no need to traverse within the RHS of a let expression.}
Edges between the nodes represent paths between expressions. The edges consist of {\em weights} (Line~\ref{alg:cfg-likelihood})
that represent the likelihood of a particular path being taken.
An edge between two nodes indicates the order of the evaluation of the program.
A node corresponding to a \il{let}-binding (Line~\ref{alg:cfg-let}) contains the bound expression
and has one outgoing edge to a node corresponding to the body expression.
A \il{case} expression (Line~\ref{alg:cfg-case}) splits the control flow $n$-ways,
where $n$ is the number of pattern matches.
{Outgoing edges of a node for a \il{case} expression have weights
  associated with them that correspond to the likelihood of taking a
  particular branch in the program.}
If available, dynamic profiling information can provide these weights.
For the purposes of this paper, however, we find it is sufficient to use trivial placeholder weights: i.e. uniform weight over all branches. 
Control flow terminates on a variable reference \systemlang{} expression.

\begin{figure}[]
  \includegraphics[scale=0.4]{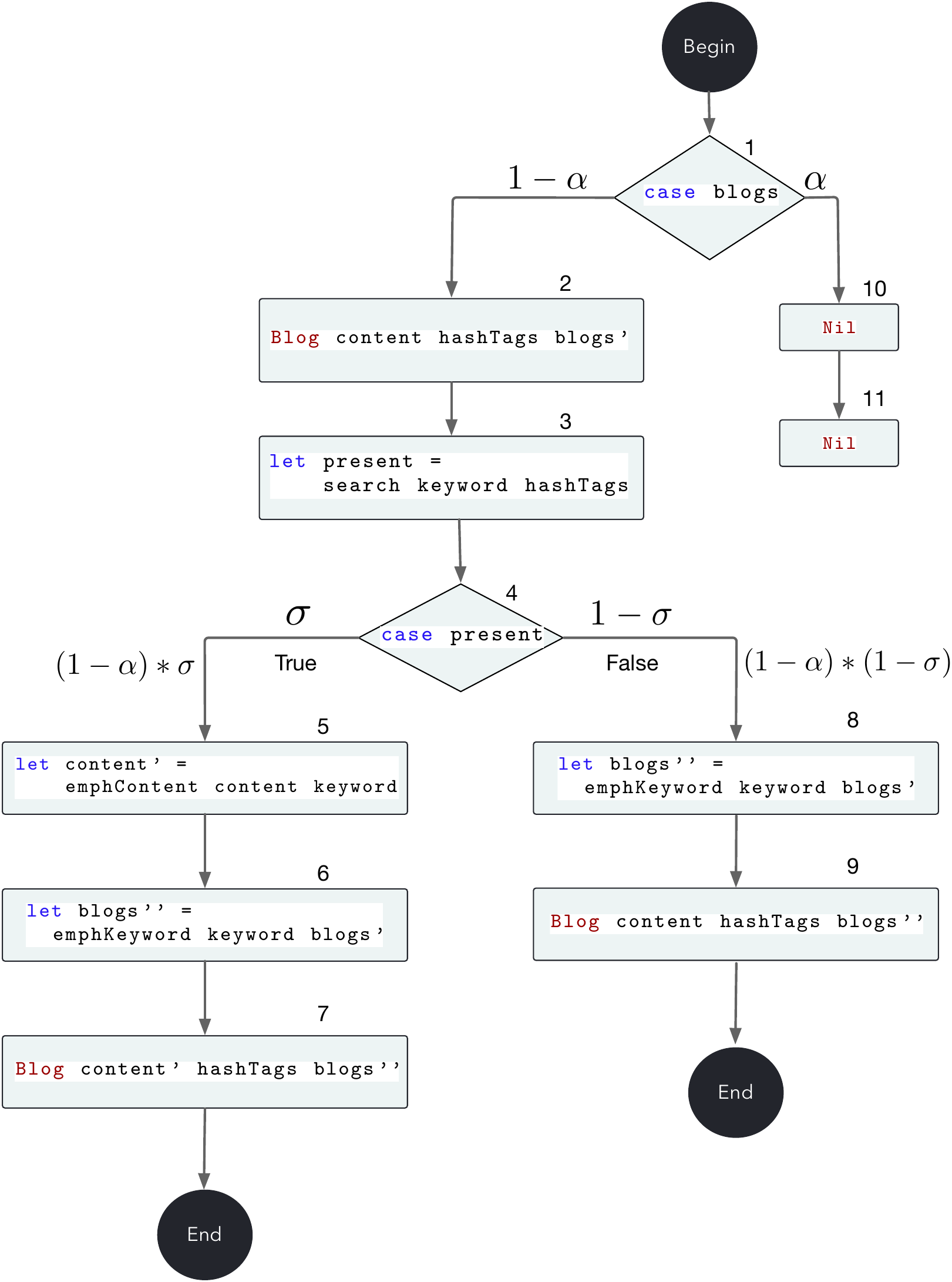}
  \caption{Control flow graph generated from function body of running example~\ref{fig:blog-traversal}.}
  \label{graph:running-cfg}
\end{figure}

\Figref{graph:running-cfg} shows the control-flow graph for the running example from \figref{fig:blog-traversal}.
{Each node corresponds to a sub-expression of the function \il{emphKeyword}.
The first \il{case} expression splits the control flow into two branches,
corresponding to whether the input list of blogs is empty or not.
The branch corresponding to the empty input list is assigned a probability $\alpha$, and thus,
the other branch is assigned a probability $1 - \alpha$.
The next node corresponds to the pattern match \il{Blog content hashTags blogs'}.
Another two-way branch follows, corresponding to whether \il{keyword} occurs in
the \il{content} of this blog or not.
We assign the probabilities $\sigma$ and $1-\sigma$ to these branches respectively.
Note that as a result, the corresponding edges in the CFG have weights $(1 - \alpha)*\sigma$ and $(1 - \alpha)*(1 - \sigma)$, as the likelihood of {\em reaching} that condition is $(1 - \alpha)$.
Each of these branches terminate by creating a new blog entry with its content potentially updated.
In the current model, $\alpha$ and $\sigma$ are 0.5: they are uniformly distributed.
}

\auditme{One intuition for why realistic branch weights are not essential to \system's optimization is that accurate weights only matter if there is a trade off between control-flow paths that are best served by different layouts. The base cases (e.g. empty list) typically contribute {\em no} ordering constraints, and in our experience, traversals tend to have a preferred order per function, rather than tradeoffs intra-function, which would reward having accurate, profile-driven branch probabilities. Hence, for now, we use uniform weights even when looking at the intra-function optimization.}
\mv{This is kinda confusing. I get the argument for why weights aren't that important to us right now, but the paper just spent some time setting up these weights just to argue that we don't need them. Could we move this discussion to future work?}
\mk{I sort of agree, but I don't want to do too much restructuring of this portion of the paper this late in the game.}

\subsection{Data Flow Analysis}
\label{design:dfa}

\vs{Section Goal: Define why we need the data flow analysis and a quick summary of how we do it in the compiler.}

We implement a straightforward analysis (use-def chain, and def-use chain) for \lstinline|let| expressions to capture dependencies between 
\lstinline|let| expressions. We use this dependence information to form dataflow edges in the field access graph (\secref{design:fieldgraph})
and to subsequently optimize the layout and code of the traversal for performance. For \textit{independent} \lstinline|let| expressions in the function we are optimizing, we can transform the function body to have these let expressions in a different order. 
(Independent implies that there are no data dependencies between such \lstinline|let| expressions. Changing the order of independent \lstinline|let| expressions will
not affect the correctness of
code, modulo exceptions.)
\rn{Nontermination should be the same irrespective of let-binding order... we could swap nontermination and a panic as long as we have a partial language though... we reserve the right to do that?}
However, we do such a transformation only when we deem it to be more cost efficient. In order to determine 
when re-ordering \lstinline|let| expressions is more cost efficient, we classify fields based on specific attributes next.

\subsection{Field Attributes For Code Motion}
\label{subsec:attributes}

\vs{Section Goal: what are these attributes and how will they help getting a lost cost ordering. }

When trying to find the best layout, we may treat the code as immutable,
but allowing ourselves to move the code around (i.e. change the order of accesses to the fields)
unlocks more possibilities for the resulting layout.
Not all code motions are valid due to existing data dependencies in the traversal.
For instance, in a sequence of two \lstinline|let| binders, the second one may reference the binding introduced in the first one: in this case, the two binders cannot be reordered.

To decide which code motions are allowed, we classify each field with one or more of the attributes: \textit{recursive, scalar, self-recursive}, or \textit{inlineable}.
Some of these attributes are derived from the ADT definition and some from the code using the ADT.
A \emph{scalar} field refers to a datatype only made up of either other primitive types, such as \lstinline|Int|.
A \emph{recursive} field refers to a datatype defined recursively.
A \emph{self-recursive} field is a recursive field that directly refers back to the datatype being defined (such as a \lstinline|List| directly referencing itself).
Finally, we call a field \emph{inlineable} if the function being optimized makes a recursive call into this field \new{(i.e.\ taking the field value as an argument)}.
Hence, an inlineable field is necessarily a recursive field.
As we show in the example below, the inlineable attribute is especially important when choosing whether or not to do code motion.

A single field can have multiple attributes.
For instance, a function (\lstinline|traverse|) doing a pre-order traversal on a \lstinline|Tree| datatype makes recursive calls on both the 
left and right children. The left and right children are recursive and self-recursive. Therefore, when looking at the scope of 
traverse, the left and right children have the attributes recursive, self-recursive and inlineable.

\ap{TODO: turn listings into figures, put them side by side}

\begin{figure}
\centering
\begin{subfigure}{.5\textwidth}
\begin{lstlisting}
foo :: List -> List 
foo lst = case lst of 
 Nil -> Nil
 Cons x rst -> 
   let x' = x ^ 100 
       rst' = foo rst 
    in Cons x' rst'
\end{lstlisting}
\caption{Function \lstinline|foo| with \lstinline|List|.}
\label{fig:foo}
\end{subfigure}\begin{subfigure}{.5\textwidth}
\centering
\begin{lstlisting}
fooFlipped :: ListFlipped -> ListFlipped 
fooFlipped lst = case lst of
 Nil' -> Nil'
 Cons' rst x -> 
   let rst' = fooFlipped rst
       x' = x ^ 100
    in Cons' rst' x' 
\end{lstlisting}
\caption{Function \lstinline|fooFlipped| with \lstinline|ListFlipped|.}
\label{fig:fooFlipped}
\end{subfigure}
\caption{Two different traversals on a list.}
\label{fig:two-diff-traversals}
\end{figure}

\paragraph{Example}
Consider the example of a list traversal shown in Figure~\ref{fig:foo}.

Here the function \lstinline|foo| does some work on the \lstinline|Int| field (it raises the \lstinline|Int| to the power of 100) and then recurs on the tail of the list.
Note that \system is compiling with dense representations, \ie{}
\lstinline|List| type is going to be 
packed in memory with a representation that stores the \lstinline|Cons| tag (one byte) followed by the \lstinline|Int| (8 bytes) followed by 
the next \lstinline|Cons| tag and so on. Hence, the \lstinline|Cons| tag and the \lstinline|Int| field are interleaved together in memory.
The function \lstinline|foo| becomes a stream processor that consumes one stream in memory and produces a dense output buffer of the same type.

On the other hand, the other way to lay out the list is to change its representation to the following. 
\begin{lstlisting}
data ListFlipped = Nil' | Cons' ListFlipped Int
\end{lstlisting}

In memory, the list has all \lstinline|Cons'| tags next to each other \new{(a unary encoding of array length!)} and the \lstinline|Int| elements all next to each other.
In such a scenario, the performance of our traversal \lstinline|foo| on the \lstinline|ListFlipped| can improve 
traversal performance due to locality when accessing elements stored side-by-side\footnote{However, this effect can disappear if the elements are very large or the amount of work done per element becomes high, such that the percent of time loading the data is amortized.}.
However, this only works if we
can subsequently change the function that traverses the list to do recursion on the tail of the list first and then call the exponentiation
function on the \lstinline|Int| field after the recursive call. If there are no data dependencies between the recursive call and the exponentiation
function, then this is straightforward. We show \lstinline|fooFlipped| with the required code motion transformation to function \lstinline|foo| accompanied
with the change in the data representation from \lstinline|List| to \lstinline|ListFlipped| as shown in Figure~\ref{fig:fooFlipped}.

\begin{table}[]
	\caption{Running times for two traversals working with their corresponding matching layouts.}
	\begin{tabular}{ccc}
	\toprule
	datatype / traversal & \lstinline|List| / \lstinline|foo| & \lstinline|List'| / \lstinline|fooFlipped| \\
	\midrule
    {1M elements} & 0.992s & 0.953s \\
	\bottomrule
	\end{tabular}
	\label{table:list-power}
\end{table}

To optimize the layout, the tail of the list is assigned the attribute of inlineable. This attribute is used by the solver to
determine a least-cost ordering to the \lstinline|List| datatype in the scope of function \lstinline|foo|. 
Whenever such code motion is possible, \system will place the inlineable field first and use code-motion to change
the body of the function to perform recursion first if data flow dependencies allow such a transformation. In fact, we made a 
similar program and ran \system on the program in the hope that it would be able to optimize the program by placing the inlineable
field first and use code-motion to change the traversal. We show the results in table~\ref{table:list-power}. We see a speedup of 4.1\% 
over our baseline datatype and traversal.

\paragraph{Structure of arrays}
The transformation of
\lstinline|List|/\lstinline|foo| to \lstinline|List'|/\lstinline|fooFlipped|
is similar to
changing the representation of the \lstinline|List| datatype to a
\textit{structure of arrays}, which causes the same types of values to be next to each other in memory.
In particular, we switch from alternating constructor tags and integer values in memory to
an array of constructor tags followed by an array of integers.

Note that the traversals \lstinline|foo| and \lstinline|fooFlipped| have access
patterns that are completely aligned with the data layout of \lstinline|List| and \lstinline|List'| respectively. The resulting speedup 
is solely a consequence of the \textit{structure of arrays} effect. This is an added benefit to the runtime in addition to ensuring that the 
access patterns of a traversal are aligned with the data layout of the datatype it traverses.

\subsection{Field Access Pattern Analysis}
\label{design:fieldgraph}

\begin{algorithm}[t]
        \caption{Recursive function for generating the field access graph}\label{alg:field-access-graph}
        \begin{algorithmic}[1]
            \small
            \Input
                \State cur: current CFG node to start processing from
                \State dcon: data constructor for which we are searching the best layout
                \State edges: field-access graph built so far
                \State lastAccessedVar: last accessed variable name
                \State dfgMap: set of data-flow edges between variables
            \EndInput
            \Output
                \State Field access graph represented as a list of edges
\EndOutput
            \Function{FieldAccessGraph}{cur, dcon, edges, lastAccessedVar, dfgMap}

                \algorithmiclet{((expr, weight), successors) = cur} \label{line-likelihood-example}
\algorithmicletmutable{lastAccessedVarMut = lastAccessedVar}
				\algorithmicletmutable{edges\textquotesingle = edges}
                \For{var : \Call{OrderedFreeVariables}{expr}} \label{line:lastAccessedVar-start}
                    \If{!\Call{BoundInPatternMatchOnDcon}{var, dcon}} \label{line:boundInDcon}
						\Continue
					\EndIf
                        \If{lastAccessedVarMut != None}
							\algorithmicmutate{edges\textquotesingle = \Call{addEdge}{edges\textquotesingle, ((lastAccessedVar, var), weight), ControlFlowTag}}  \label{line:cfg-edge}
                            \If{\Call{lookup}{(lastAccessedVar, var), dfgMap}}
                                \algorithmicmutate{edges\textquotesingle = \Call{addEdge}{edges\textquotesingle, ((lastAccessedVar, var), weight), DataFlowTag}}  \label{line:dfg-edge}
                            \EndIf
\EndIf
                        \algorithmicmutate{lastAccessedVarMut = var}
                \EndFor \label{line:lastAccessedVar-end}
                
                \For{succ : successors} \label{line:succ-rec-start}
                    \algorithmiclet{edges\textquotesingle\textquotesingle = \Call{FieldAccessGraph}{succ, dcon, edges\textquotesingle, lastAccessedVarMut, dfgMap}}
                    \algorithmicmutate{edges\textquotesingle = \Call{merge}{edges\textquotesingle, edges\textquotesingle\textquotesingle}}
                \EndFor \label{line:succ-rec-end}

                \State \Return edges\textquotesingle

            \EndFunction

\end{algorithmic}
    \end{algorithm}

\vs{Goal: Define how the access patters are derived from the control flow graph of a traversal. How its incoded in a graph. }

After constructing the \CFG and \DFG for a function definition, we utilize them to inspect the type
of each input parameter---one data constructor at a time---and construct
a \emph{field-access graph} for it.
Algorithm~\ref{alg:field-access-graph} shows the psuedocode for generating the field-access graph. 
This graph represents the temporal ordering of accesses among its fields.

\begin{figure}[]
  \includegraphics[scale=0.45]{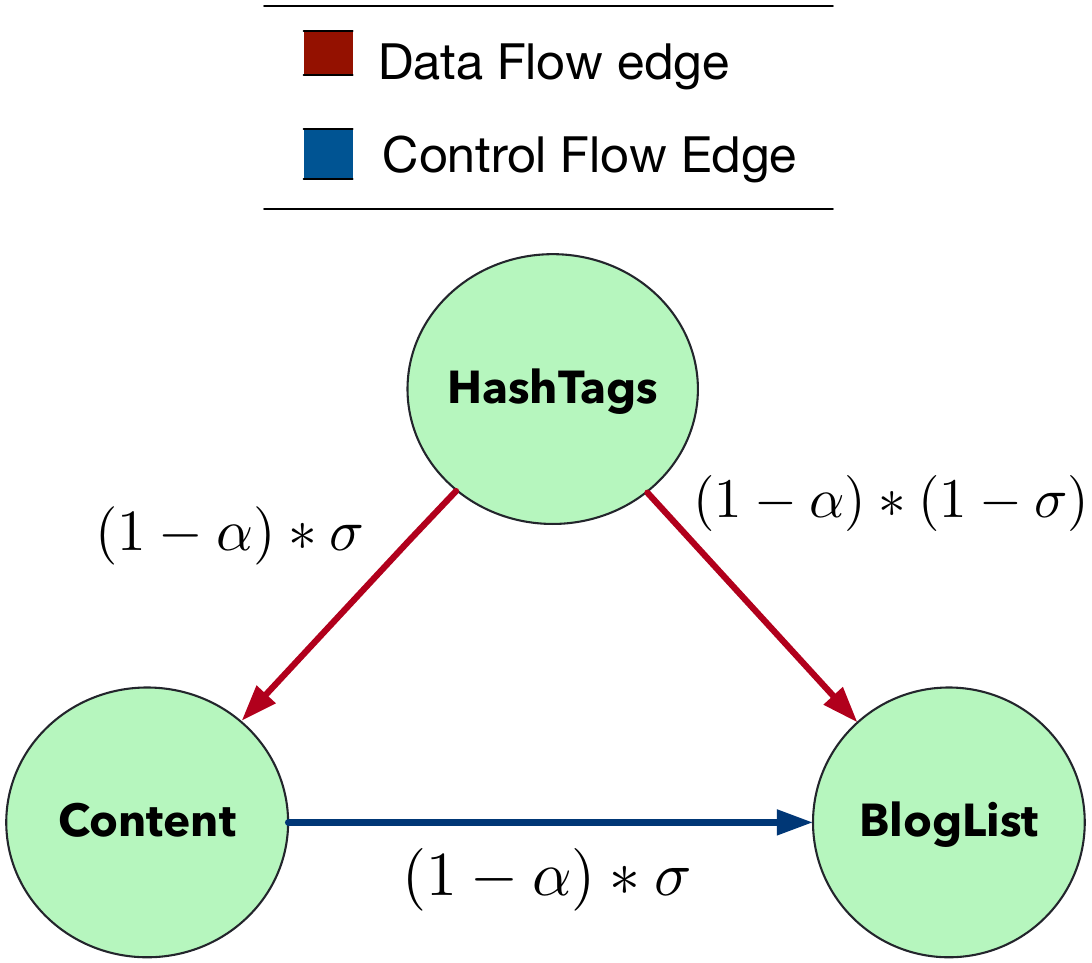}
  \caption{Field access graph $G$ generated from \CFG of running
    example~\ref{graph:running-cfg}.  Here we use field types to
    uniquely name them, since Haskell record syntax was not used in
    this example to give each field a name.}
  \label{graph:running-access}
\end{figure}

The fields of the data constructor form the nodes of this graph.
A directed edge from field $f_i$ to field $f_j$ is added if $f_i$ is accessed immediately before $f_j$. Lines~\ref{line:lastAccessedVar-start} to ~\ref{line:lastAccessedVar-end}
in Algorithm~\ref{alg:field-access-graph} show how we keep track of the last accessed field and form an edge if possible. 
A directed edge can be of two different types. In addition, each edge has a associated weight which indicates the likelihood of accessing $f_i$ before $f_j$, which is computed using the \CFG.
An edge can either be a data flow edge or a control flow edge (Lines~\ref{line:cfg-edge} and ~\ref{line:dfg-edge}). In \Figref{graph:running-access}, the red edge is a data flow edge and the blue edge is a control flow edge.

\textit{Data Flow Edge} indicates an access resulting from a data flow dependence between the fields $f_i$ and $f_j$. In our source language, a data flow edge is induced by
a \lstinline|case| expression.
A data flow edge implies that the code that represents the access is rigid in structure and changing it can make our transformation invalid.

\textit{Control Flow Edge} indicates an access that is not data-flow dependent. It is caused by the control flow of the program. Such an edge does not induce strict constraints on the code that induces the edge. The code is malleable in case of such accesses. This gives way to an optimization search space via code motion of \lstinline|let| expressions. The optimization search space involves transformation of the source code, i.e, changing the access patterns at the source code level.

The field-access graph $G$ is a directed graph, which consists of edges of the two types between fields of a datatype and can have cycles. The directed nature of the edges enforces a temporal relation between the corresponding fields. More concretely, assume that an edge $e$ that connects two vertices representing fields $\fields_{a}$ (source of $e$) and $\fields_{b}$ (target of $e$). We interpret $e$ as an evidence that field $\fields_{a}$ is accessed before field $\fields_{b}$. The weight $w$ for the edge $e$ is the probability that this access will happen based on statically analyzing a function. 

In our analysis, for a unique path through the traversal, we only account for the {\em first} access to any two fields. If two fields are accessed in a different order later on, the assumption is that the start address of the fields is likely to be in cache and hence it does not incur an expensive fetch call to memory. In fact, we tested our hypothesis by artificially making an example where say field $f_{a}$ is accessed first, field $f_{b}$ is accessed after $f_{a}$ after which we constructed multiple artificial access edges from $f_{b}$ to $f_{a}$, which might seem to suggest placing $f_{b}$ before $f_{a}$. However, once the cache got warmed up and the start addresses of $f_{a}$ and $f_{b}$ are already in cache, the layout did not matter as much. This suggests that prioritizing for the first access edge between two unique fields along a unique path is sufficient for our analysis.

Two fields can be accessed in a different order along different paths through a traversal. This results in two edges between the fields. (The edges are in reversed order.) We allow at most two edges between any two vertices with the constraint that they have to be in the opposite direction and come from different paths in the traversal. If two fields are accessed in the same order along different paths in the traversal, we simply add the probabilities and merge the edges since they are in the same direction.

In order to construct $G$, we topologically sort the control flow graph of a function and traverse it in the depth-first fashion by doing recursion on the successors of the current cfg node (Lines~\ref{line:succ-rec-start} to~\ref{line:succ-rec-end}).
As shown in line~\ref{line:boundInDcon}, we check if a variable in an alias to a field in the data
constructor for which we are constructing the field-access graph $G$.
As we process each node (i.e. a primitive expression such as a single
function call), we update the graph for any direct or indirect
references to input fields that we can detect.
We ignore new variable bindings that refer to newly allocated rather
than input data---they are not tracked in the access graph.
\auditme{We traverse the control-flow graph once, but we maintain the
  last-accessed information at each CFG node, so when we process a field
  access at an expression, we consult what was previously-accessed at
  the unique precedecessor of the current CFG node.}
\Figref{graph:running-access} shows the generated access graph from the control flow graph in \figref{graph:running-cfg}. It also shows the probability along each edge obtained from the control flow graph.
\rn{Ideally I'd love to see this algorithm in psuedocode, and the text focus on the running example rather than walking through the algorithm in english.}
\vs{Yeah, psuedocode would be nice, I don't think there is time now, but definitely going to add to TODO list.}

As we are traversing the nodes of the control flow graph and generating directed edges in $G$, we use the likelihood of accessing that cfg node as the weight parameter (Line~\ref{line-likelihood-example}).

\subsection{Finding a Layout}
\label{design:genconstraints}

We use the field-access graph $G$ to encode the problem of finding a better layout as an Integer Linear Program (ILP). Solving the problem yields a cost-optimal field order for the given pair of a data constructor and a function.

\subsubsection{ILP Constraints}
In our encoding, each field in the data constructor is represented by a variable, $f_0, f_1, \dots$ As a part of the result, each variable will be assigned a unique integer in the interval $[0, n -1]$, where $n$ is the number of fields. Intuitively, each variable represents an index in the sequence of fields.

The ILP uses several forms of constraints, including two forms of \emph{hard} constraints:
\begin{gather}
  \forall_{0 \leq i < n}  \qquad 0 \leqslant f_{i}  < n \label{eq:valid}\\
  \forall_{0 \leq i < j < n} \qquad f_i \neq f_j   \label{eq:uniq}
\end{gather}
The constraints of form~\ref{eq:valid} ensure that each field is mapped to a valid index, while the constraints of form~\ref{eq:uniq} ensure that each field has a unique index.
Constraints of either form must hold because each field must be in a valid location.

Hard constraints define valid field orderings but not all such reorderings improve efficiency, \system's main goal.
To fulfil the goal, beside the hard constraints we introduce {\em soft} ones. Soft constraints come from the field access analysis. For example, assume that based on the access pattern of a function, we would {\em prefer} that field $a$ goes before field $b$. We turn such a wish into a constraint. If the constraint cannot be satisfied, it will not break the correctness. In other words, such constraints can be broken, and that is why we call them soft.

\subsubsection{Cost Model}
\system encodes these soft constraints in the form of an abstract cost model that assigns a cost to a given layout (assignment of fields to positions) based on how efficient it is expected to be given the field-access graph.

To understand the intuition behind the cost model, note that the existence of an edge from field $f_i$ to field $f_j$ in the field-access graph means that there exists at least one path in the control-flow graph where $f_i$ is accessed and $f_j$ is the next field of the data constructor that is accessed.
In other words, the existence of such an edge implies a preference {\em for that control flow path} for field $f_i$ to be immediately before field $f_j$ in the layout so that the program can continue a linear scan through the packed buffer. 
Failing that, it would be preferable for $f_j$ to be ``ahead'' of $f_i$ in the layout so the program does not have to backtrack in the buffer.
We can thus consider the costs of the different layout possibilities of $f_i$ and $f_j$:

\begin{description}
	\item[$C_{\mathit{succ}}$] ($f_j$ immediately after $f_i$): This is the best case scenario: the program traverses $f_i$ and then uses $f_j$.
	\item[$C_{\mathit{after}}$] ($f_j$ after $f_i$ in the buffer): If $f_j$ is after $f_i$, but not {\em immediately} after, then the code can proceed without backtracking through the buffer, but the intervening data means that either a shortcut pointer or a \new{extra traversal} must be used to reach $f_j$, adding overhead.
	\item[$C_{\mathit{pred}}$] ($f_j$ immediately before $f_i$): Here, $f_j$ is {\em earlier} than $f_i$ in the buffer. Thus, the program will have already skipped past $f_j$, and some backtracking will be necessary to reach it. This incurs {\em two} sources of overhead: skipping past $f_j$ in the first place, and then backtracking to reach it again.
	\item[$C_{\mathit{before}}$] ($f_j$ before $f_i$ in the buffer): If, instead, $f_j$ is farther back in the buffer than $f_i$, then the cost of skipping back and forth is greater: in addition to the costs of pointer dereferencing, because the fields are far apart in the buffer, it is less likely $f_j$ will have remained in cache (due to poorer spatial locality).
\end{description}

We note a few things. 
First, the {\em exact} values of each of these costs are hard to predict. 
The exact penalty a program would pay for jumping ahead or backtracking depends on a variety of factors such as cache sizes, number of registers, cache line sizes, etc.

However, we use our best intuition to statically predict these costs based on the previously generated access graph.
Note the existence of two types of edges in our access graph. An edge can either be a data flow edge or a control flow edge.
For a data flow edge, the code is rigid. Hence the only axis we have available for transformation is the datatype itself. For a data flow edge, the costs are showed in Eq~\ref{eq:costs-strong-edge}. Here, we must respect the access patterns in the original code which lead to the costs in Eq~\ref{eq:costs-strong-edge}.
\begin{equation}
C_{\mathit{succ}} < C_\mathit{after} < C_\mathit{pred} < C_\mathit{before}
\label{eq:costs-strong-edge}
\end{equation}

Note that a control flow edge signifies that the direction of access for an edge is transformable. We could reverse the access in the code without breaking the correctness of the code. We need to make a more \textit{fine-grained} choice.
This choice involves looking at the attributes of the fields and making a judgement about the costs given we know the attributes of the fields. As shown in Sec~\ref{subsec:attributes} we would like to have the field with an \textit{inlineable} attribute placed first.
Hence, in our cost model, if $f_i$ is inlineable, then we follow the same costs in Eq~\ref{eq:costs-strong-edge}. However, if $f_j$ is inlineable and $f_i$ is not, we would like $f_j$ to be placed before $f_i$. For such a layout to endure, the costs should change to Eq~\ref{eq:costs-inlineable}.
For other permutations of the attributes, we use costs that prioritize placing the inlineable field/s first. 
\begin{equation}
C_\mathit{pred} < C_\mathit{before} < C_{\mathit{succ}} < C_\mathit{after}
\label{eq:costs-inlineable}
\end{equation}

\subsubsection{Assigning Costs to Edges}

\system uses the field-access graph and the cost model to construct an objective function for the ILP problem.
Each edge in the access graph represents one pair of field accesses with a preferred order. Thus, for each edge $e = (i, j)$, \system can use the indices of the fields $f_i$ and $f_j$ to assign a cost, $c_e$, to that pair of accesses following the rules below.
\begin{table}[h]
\centering
\begin{tabular}{ll}
If $f_j$ is right after $f_i$, then assign cost $C_\mathit{succ}$, i.e.:
&
$
(f_j - f_i) = \phantom{-}1 \implies c_e = C_\mathit{succ}
$.
\\
If $f_j$ is farther ahead of $f_i$, then assign cost $C_\mathit{after}$, i.e.:
&
$
(f_j - f_i) > \phantom{-}1 \implies c_e = C_\mathit{after}
$.
\\
If $f_j$ is immediately {\em before} $f_i$, then assign cost $C_\mathit{pred}$, i.e.:
&
$
(f_j - f_i) = -1 \implies c_e = C_\mathit{pred}
$.
\\
And if $f_j$ is farther before $f_i$, then assign cost $C_\mathit{before}$, i.e.:
&
$
(f_j - f_i) < -1 \implies c_e = C_\mathit{before}
$.
\end{tabular}
\end{table}

The cost of each edge, $c_e$ must be multiplied by the {\em likelihood} of that edge being exercised, $p_e$, which is also captured by edge weights in the field-access graph. Combining these gives us a total estimated cost for any particular field layout:
\begin{equation}
C = \sum_{e \in E} c_e \cdot p_e
\label{eq:objective}
\end{equation}
This is the cost that our ILP attempts to minimize, subject to the hard constraints~\ref{eq:valid} and~\ref{eq:uniq}.

\subsubsection{Greedy layout ordering}

Since there is a tradeoff in finding optimal layout orders and compile times when using a solver based optimization. We implemented a greedy algorithm that traverses the access graph for a data constructor
in a greedy fashion. It starts from the root node of the access graph which is the first field in the layout of the data constructor and greedily (based on the edge weight) visits the next field. 
We fix the edge order for a control flow edge as the original order and do not look at field attributes. However, after the greedy algorithm picks a layout we match the let expressions to the layout order to make sure the code matches the layout order. A greedy algorithm is potentially sub-optimal when it comes to finding the best performing layout; however, the compile time is fast.

\subsection{Finding a global layout}
\label{design:global}

A data constructor can be used across multiple functions, therefore, we need to find a layout order that is optimal globally.
To do so, we take constraints for each function and data constructor pair and combine them uniformly, that is, a uniform weight for each function.
We then feed the combined constraints to the solver to get a globally optimal layout for that data constructor. The global optimization finds a globally optimal layout for all data constructors in the program. Once the new global layout is chosen for a data constructor, we re-write the 
entire program such that each data constructor uses the optimized order of fields.
In the evaluation, we use the global optimization. However, we only show the data constructor that constitutes the major part of the program.

\subsection{Finding a layout for functions with conflicting access patterns}
\label{design:conflicting-access-patterns}
Consider the datatype definition \lstinline|D| with two fields \lstinline|A, B|:
\begin{lstlisting}
    data D = D A B
\end{lstlisting}
We have conflicting access patterns for \lstinline|D|
if there are two functions \lstinline|F1| and \lstinline|F2| that access the fields of \lstinline|D| in two opposite orders.
Assume \lstinline|F1| accesses \lstinline|A| first and then \lstinline|B|, while \lstinline|F2|
accesses \lstinline|B| first and then \lstinline|A|. After combining edges across the two functions, we will have two edges that
are in opposition to each other. Since we give a uniform weight to each function, the edges will also have a uniform weight.
In this case,
placing \lstinline|A| before \lstinline|B| or \lstinline|B| before \lstinline|A| are equally acceptable in our model, and
\system defers to the solver to get one of the two possible layouts.

With the two equally good layouts, \system's default solver-based mode chooses the layout favoring the function it picked first. For instance, if \lstinline|F1| is defined earlier in the program, the order favoring \lstinline|F1| will be picked.
In the greedy-heuristic mode, since both \lstinline|A| and \lstinline|B| are root nodes, \system will pick the
first root node in the list of root nodes and start greedily visiting child nodes from that particular root node, thus, fixing a layout order.
If we determined that \lstinline|F2| has a greater weight than \lstinline|F1|, then the layout preferred by
\lstinline|F2| would be chosen.

\section{Implementation}\label{sec:impl}

\begin{figure}[]
	\includegraphics[scale=0.4]{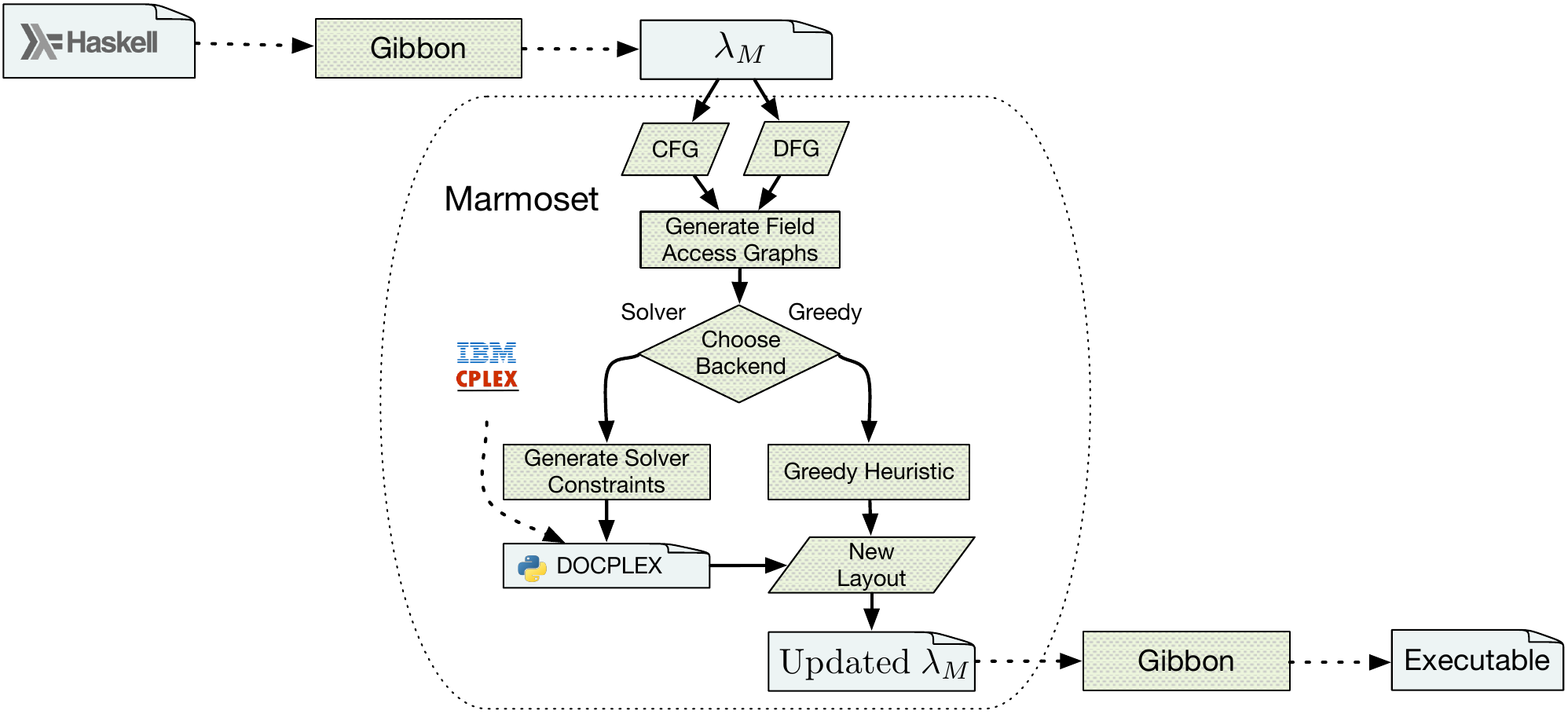}
	\caption{The overall pipeline of \system.}
	\label{figure:overall-pipeline}
\end{figure}

We implement \system{} in the open-source Gibbon compiler\footnote{https://github.com/iu-parfunc/gibbon/}.
Figure~\ref{figure:overall-pipeline} gives an overview of the overall pipeline. 
Gibbon is a whole-program micropass compiler that compiles a polymorphic,
higher-order subset of (strict) Haskell.

The Gibbon front-end uses standard whole-program compilation and monomorphization techniques~\cite{urweb-icfp}
to lower input programs into a first-order, monomorphic IR (\systemlang{}).
Gibbon performs location inference on this IR to convert it into a LoCal program,
which has regions and locations, essentially, buffers and pointer arithmetic.
Then a big middle section of the compiler is a series of LoCal $\rightarrow$ LoCal compiler passes that
perform various transformations.
Finally, it generates C code.\csk{perhaps this paragraph is unnecessary, but we don't say much about gibbon anywhere else.}
Our extension operates towards the front-end of the compiler, on \systemlang{}.
We closely follow the design described in Section~\ref{sec:design} to construct the
control-flow graph and field-access graph, and use the standard Haskell graph library\footnote{https://hackage.haskell.org/package/containers} in our implementation.

To solve the constraints, we use IBM's DOCPLEX (Decision Optimization CPLEX),
because its API allows high level modelling such as logical expressions like implications, negations, logical AND etc. with relatively low overhead. 
Unfortunately, there isn't a readily available Haskell library that can interface with DOCPLEX.
Thus, we use it via its library bindings available for Python.
Specifically, we generate a Python program that feeds the constraints to DOCPLEX and
outputs an optimum field ordering to the standard output,
which \system{} reads and parses, and then reorders the fields accordingly.

 \section{Evaluation}

We evaluate \system on three applications.
First is a pair of microbenchmarks: a list length function
(Section~\ref{subsubsec:length}) and a logical expression evaluator (Section~\ref{subsubsec:logic}).
Second is a small library of operations with binary trees (Section~\ref{subsec:trees}).
Third is a blog management software based on the \lstinline|BlogList| example from the
Sections~\ref{sec:overview}--\ref{sec:design} (Section~\ref{subsect:blog}).
Besides the run times, we take a closer look at how \system affects
cache behavior (Section~\ref{subsec:cache}) and compile times (Section~\ref{subsec:heuristic-tradeoffs}).

We detail the impact of various datatype layouts on the performance.
As the baseline, we use \gibbon, the most closely related prior work.
We also compare \system with \mlton (Section~\ref{subsec:mlton}).
For each benchmark, we run 99 iterations and report the run-time mean, median and
the 95\% confidence interval.

\subsection{Experimental Setup}
We run our benchmarks on a server-class machine with
64 CPUs, each with two threads. The CPU model is AMD Ryzen Threadripper 3990X with 2.2 GHz clock speed. The L1 cache size is 32 KB, L2 cache size is 512 KB and L3 cache size is 16 MB.
We use \gibbon's default C backend and call GCC 10.2.0 with \texttt{-O3} to generate binaries.

\subsection{Micro Benchmarks}

We use two microbenchmarks to explore specific performance penalties imposed by a sub-optimal data layout.

\subsubsection{{\bf Computing the length of a linked-list}}\label{subsubsec:length}

\new{
This benchmark demonstrates the cost of de-referencing memory addresses which are not present in the cache.
It uses the linked list datatype:
}
\begin{lstlisting}
data List = Nil | Cons Content List 
\end{lstlisting}
\new{
If each element of the list is constructed using \lstinline{Cons},
the traversal has to de-reference a pointer---to {\em jump over} the content---each time to access the tail of the list.
This is an expensive operation, especially if the target memory address is not present in the cache.
In contrast, if the \lstinline{Content} and \lstinline{List} fields were swapped,
then to compute the length, the program only has to traverse $n$ bytes for a list of length $n$---one byte
per \lstinline{Cons} tag---which is extremely efficient.
Essentially, \system transforms program to use the
following datatype, while preserving its behavior:}

\begin{lstlisting}
data List' = Nil' | Cons' List' Content
\end{lstlisting}

In our experiment, the linked list is made of 3M elements and each element contains some
content (a \textit{Pandoc} \lstinline|Inline| datatype) that occupies roughly 5KB.
\csk{wanted to say something like: 5KB is much bigger than a cache line which makes the pointer de-refrence expensive for \lstinline{Cons}-list and having 3M elements ensures that even the \lstinline{Snoc} tags overflow, so it's not like Snoc cheats by fitting everything in a cache line.}
As seen in table~\ref{table:Measure-Length}\footnote{The performance of \lstinline|List'| layout compiled through \gibbon differs from the greedy and solver version as \system does code motion to reorder let expressions which results in different code.}, the performance of the list constructed using the original \lstinline{List} is $\sim$60$\times$ worse than the performance with the \system-optimized, flipped layout.
Not only does \lstinline{List} have poor data locality and cache behavior,
but it also has to execute more instructions to de-reference the pointer.
Both \systemGreedy and \systemSolver choose the flipped layout \lstinline|List'|.

\begin{table}[]
	\setlength{\tabcolsep}{2pt} \centering
	\ra{1}
	\caption{List length run times (99 runs): Mean, Median and 95\% confidence interval (ub, lb) in milliseconds (ms) for different
	layouts when compiled with \gibbon. The last two columns show the runtime for the layout chosen by
	\systemGreedy and \systemSolver.
	The numbers in \textcolor{blue}{Blue} correspond to the lowest running time and the numbers in \textcolor{red}{Red} correspond to the highest running time.
	}
	\begin{tabular}{@{}p{0.6in}rcrcrcrcrcrcrcr@{}}\toprule
	 & \multicolumn{3}{|c|}{{\gibbon}} &
	 & \multicolumn{3}{c|}{{\system}}\\ \midrule
		
	 & \multicolumn{1}{c}{{List}} &
	 & \multicolumn{1}{c}{{List'}} &
	 & \multicolumn{1}{c}{\small{\systemGreedy}} &
	 & \multicolumn{1}{c}{\small{\systemSolver}}\\ \midrule
	Mean
	         & \textcolor{red}{60.731}&&  {1.052}&&  \textcolor{blue}{0.981} &&  {0.994} \\
	Median
	         &  \textcolor{red}{60.730}&&   {1.017}&&   \textcolor{blue}{0.978} &&   {0.979}\\
ub
	               &  {60.737}&&   {1.078}&&   {0.990}&&   {1.022}\\
	lb
	               &  {60.725}&&   {1.025}&&   {0.973} &&   {0.964}\\
	\bottomrule
	\end{tabular}
	\label{table:Measure-Length}
\end{table}

\subsubsection{{\bf Short-circuiting logical And/Or operators}}\label{subsubsec:logic}

\new{
These microbenchmarks operate over a synthetically generated, balanced syntax-tree with a height of 30.
The intermediate nodes can be one of \lstinline{Not}, \lstinline{Or}, or \lstinline{And}, selected at random, and
the leaves consist of booleans wrapped in a \lstinline{Val}.
The evaluator function evaluates the expression tree and performs short circuiting
for expressions like \lstinline{and} and \lstinline{or} in the expression tree. We measure the performance 
of this evaluator function by changing the order of the left and right subtrees. 
}
The syntax-tree datatype is defined as follows:
\begin{lstlisting}
data Exp = Val Bool | Not Exp | Or Exp Exp | And Exp Exp
\end{lstlisting}
Since the short circuiting evaluates from left to right order of the \lstinline{Exp}, changing the order of the left and right subtrees would affect the performance of the traversal. As can be seen in table~\ref{table:Logical-Expression-Evaluator}, 
the layout where the left subtree is serialized before the right subtree results in better performance compared to the tree where the the right subtree is serialized before the left one. This is as expected since in the latter case, the traversal has to jump 
over the right subtree serialized before the left one in order to evaluate it first and then depending on the result of the left subtree possibly jump back to evaluate the right subtree. This results in poor spatial locality and hence worse performance. \systemGreedy and \systemSolver are able to identify the layout transformations that would give the best performance, which matches the case where the left subtree is serialized before the right subtree.
Table~\ref{table:Logical-Expression-Evaluator}\footnote{The layout chosen by \systemGreedy and \systemSolver
is same as lr, the performance differs from the lr layout compiled through \gibbon as \system does code motion 
which results in different code.} shows the performance results. 

\begin{table}[]
	\setlength{\tabcolsep}{2pt} \centering
	\ra{1}
	\caption{Logical expression evaluator run times in seconds (s).
	Legend: l - left subtree, r - right subtree  of the decision tree.}
	\begin{tabular}{@{}p{0.6in}rcrcrcrcrcrcrcr@{}}\toprule
	 & \multicolumn{3}{|c|}{{\gibbon}} &
	 & \multicolumn{3}{c|}{{\system}}\\ \midrule

	 & \multicolumn{1}{c}{{lr}} & 
	 & \multicolumn{1}{c}{{rl}} &
	 & \multicolumn{1}{c}{\small{\systemGreedy}} &
	 & \multicolumn{1}{c}{\small{\systemSolver}}\\ \midrule
	Mean
	         &   \fnumthree{4.372658620370371}
			 &&   \textcolor{red}{\fnumthree{6.669544222222224}}
			 &&   \fnumthree{3.600981037037037}
			 &&   \textcolor{blue}{\fnumthree{3.594482435185185}}

			\\
	Median
	        &  \fnumthree{4.313438}
			&&   \textcolor{red}{\fnumthree{6.625107}}
			&&    \textcolor{blue}{\fnumthree{3.534958}}
			&&  {\fnumthree{3.5430935000000003}}

			\\
ub          
	               &    \fnumthree{4.403732645795353}
				   &&    \fnumthree{6.6931755244089395}
				   &&    \fnumthree{3.648194996569051}
				   &&    \fnumthree{3.635610184890118}

				   \\
	lb         
	               &    \fnumthree{4.341584594945388}
				   &&    \fnumthree{6.645912920035508}
				   &&    \fnumthree{3.553767077505023}
				   &&    \fnumthree{3.553354685480252}
				   \\
	\bottomrule
	\end{tabular}
	\label{table:Logical-Expression-Evaluator}
\end{table}

\subsection{Binary Tree benchmarks}\label{subsec:trees}

We evaluate \system on a few binary tree benchmarks. These include adding one to all values stored in the internal nodes of the tree, exponentiation on integers stored in internal nodes, copying a tree and getting the right-most leaf value in the tree.
For the first three benchmarks, the tree representation we use is:
\begin{lstlisting}[]
data Tree = Leaf | Node Int Tree Tree
\end{lstlisting}
For right-most, the tree representation we use is:
\begin{lstlisting}[]
data Tree = Leaf Int | Node Tree Tree
\end{lstlisting}

\subsubsection{Add one tree}

This benchmark takes a full binary tree and increments the values stored in the internal nodes of the tree. 
We show the performance of an aligned preorder, inorder and postorder traversal in addition to a misaligned 
preorder and postorder traversal of the tree. Aligned traversals are ones where the data representation exactly
matches the traversal order, for instance, a preorder traversal on a preorder representation of the tree. 
A misaligned traversal order is where the access patterns of the traversal don't match the data layout of the tree.
For instance, a postorder traversal on a tree serialized in preorder. 
Table~\ref{table:Add-one-Tree} shows the performance numbers. We show also the performance of \systemGreedy and \systemSolver.
\systemSolver picks the aligned postorder traversal order which is best performing. It makes the recursive calls to the left and right children of the
tree first and increments the values stored in the internal nodes once the recursive calls return. The tree representation is also changed to a postorder
representation with the \lstinline|Int| placed after the left and right children of the tree. 
This is in part due to the structure of arrays effect, as the \lstinline|Int| are placed closer to each other.
\systemGreedy on the other hand picks the 
aligned preorder traversal because of its greedy strategy which prioritizes placing the \lstinline|Int| before the left and right subtree.
The tree depth was set to 27. At this input size, the \misalignedPost traversal failed due to memory errors.
We see significant slowdowns with \misalignedPre, $\sim$35$\times$ slower, because of the skewed access patterns of the traversal.

\begin{table}[]
    \small
    \setlength{\tabcolsep}{3.2pt} \centering
    \ra{1}
    \caption
            { Binary Tree benchmarks runtime statistics in seconds (s).
              \misalignedPre -- A post order traversal on a pre order layout of the tree.
              \misalignedPost -- A pre order traversal on a post order layout of the tree.
              \alignedPre -- A pre order traversal on a pre order layout of the tree.
              \alignedIn -- An in order traversal on an in order layout of the tree.
              \alignedPost -- A post order traversal on a post order layout of the tree.
              }
    \begin{tabular}{@{}p{0.6in}rcrcrcrcrcrcrcrcr@{}}\toprule
        & \multicolumn{12}{|c|}{{\gibbon}} &
        & \multicolumn{3}{c|}{{\system}}\\ \midrule

     &
     &
     &
     & \multicolumn{1}{c}{\scriptsize{\misalignedPre}} &
     & \multicolumn{1}{c}{\scriptsize{\misalignedPost}} &
     & \multicolumn{1}{c}{\scriptsize{\alignedPre}} &
     & \multicolumn{1}{c}{\scriptsize{\alignedIn}} &
     & \multicolumn{1}{c}{\scriptsize{\alignedPost}} &
     & \multicolumn{1}{c}{\scriptsize{\systemGreedy}} & 
     & \multicolumn{1}{c}{\scriptsize{\systemSolver}}\\ \midrule
    Add\:One\:Tree\\
    Mean
        &
        &
        &&   \textcolor{red}{{45.506}}
        &&   memory error
        &&   \snumfour{1.3188707407407407}
        &&   \snumfour{1.3274673981481484}
        &&   \snumfour{1.2906462222222224}
        &&   \snumfour{1.3189152222222222}
        &&   \textcolor{blue}{\snumfour{1.2789202828282824}}
        \\
    Median
        &
        &
        &&   \textcolor{red}{{45.507}}
        &&   memory error
        &&   \snumfour{1.2915475}
        &&   \snumfour{1.301803}
        &&   \snumfour{1.288307}
        &&   \snumfour{1.2912845}
        &&   \textcolor{blue}{\snumfour{1.275409}}
        \\
    ub
        &
        &
        &&    {45.661}
        &&    memory error
        &&    \snumfour{1.334508271008376}
        &&    \snumfour{1.342853017459503}
        &&    \snumfour{1.2961216110913416}
        &&    \snumfour{1.3346949244887727}
        &&    \snumfour{1.2848468714070156}
        \\
    lb
                    &
                    &
                    &&    {45.351}
                    &&    memory error
                   &&    \snumfour{1.3032332104731055}
                   &&    \snumfour{1.3120817788367938}
                   &&    \snumfour{1.2851708333531031}
                   &&    \snumfour{1.3031355199556718}
                   &&    \snumfour{1.2729936942495492}
      \\
      \\
    Exponentiation\:Tree\\
    Mean
             &
             &
             &&    \textcolor{red}{{45.520}}
             &&    memory error
             &&   \snumfour{1.3388337314814818}
             &&   \snumfour{1.340895601851852}
             &&   \snumfour{1.2963395151515151}
             &&   \snumfour{1.3362375370370372}
             &&   \textcolor{blue}{\snumfour{1.294804525252525}}

            \\
    Median
             &
             &
             &&    \textcolor{red}{{45.567}}
             &&    memory error
            &&   \snumfour{1.3129675}
            &&   \snumfour{1.3147785}
            &&   \snumfour{1.293376}
            &&   \snumfour{1.310292}
            &&   \textcolor{blue}{\snumfour{1.291829}}
            \\
    ub
                    &
                    &
                    &&    {45.702}
                    &&    memory error
                   &&    \snumfour{1.3543494934201488}
                   &&    \snumfour{1.356565502024642}
                   &&    \snumfour{1.3019615832301774}
                   &&    \snumfour{1.3521300688353535}
                   &&    \snumfour{1.3006312577538843}

                   \\
    lb
                    &
                    &
                    &&    {45.338}
                    &&    memory error
                   &&    \snumfour{1.3233179695428148}
                   &&    \snumfour{1.325225701679062}
                   &&    \snumfour{1.2907174470728529}
                   &&    \snumfour{1.320345005238721}
                   &&    \snumfour{1.2889777927511659}

                   \\
      \\
    Copy\:Tree\\
    Mean
             &
             &
             &&    \textcolor{red}{{45.524}}
             &&    memory error
             &&   \snumfour{1.312177046296296}
             &&   \snumfour{1.3230825185185187}
             &&   \snumfour{1.2816441919191919}
             &&   \snumfour{1.3116497129629627}
             &&   \textcolor{blue}{\snumfour{1.2786672424242425}}
            \\
    Median
             &
             &
             &&    \textcolor{red}{{45.529}}
             &&    memory error
            &&   \snumfour{1.2860195}
            &&   \snumfour{1.2969865}
            &&   \snumfour{1.278153}
            &&   \snumfour{1.284783}
            &&   \textcolor{blue}{\snumfour{1.27603}}
            \\
    ub
                    &
                    &
                    &&    {45.673}
                    &&    memory error
                   &&    \snumfour{1.327728808958863}
                   &&    \snumfour{1.33875068512881}
                   &&    \snumfour{1.2873179678481805}
                   &&    \snumfour{1.3274813171692397}
                   &&    \snumfour{1.2843587943086343}
                   \\
    lb
                    &
                    &
                    &&    {45.374}
                    &&    memory error
                   &&    \snumfour{1.296625283633729}
                   &&    \snumfour{1.3074143519082273}
                   &&    \snumfour{1.2759704159902032}
                   &&    \snumfour{1.2958181087566858}
                   &&    \snumfour{1.2729756905398506}
                   \\
    \bottomrule
    \end{tabular}
    \label{table:Add-one-Tree}
    \label{table:Exponentiation-Tree}
	\label{table:Copy-Tree}
\end{table} 
\subsubsection{Exponentiation Tree}

This traversal does exponentiation on the values stored in the internal nodes of the tree. It is more computationally intensive than incrementing the value.
We raise the \lstinline|Int| to a power of 10 on a tree of depth 27. Table~\ref{table:Exponentiation-Tree} shows the performance of the different layout and traversal orders.
\systemSolver picks the \alignedPost representation which is the best performing, whereas \systemGreedy picks the \alignedPre representation. 

\subsubsection{Copy Tree}

Copy-tree takes a full binary tree and makes a fresh copy of the tree in a new memory location.
We use a tree of depth 27 in our evaluation. Table~\ref{table:Copy-Tree} shows the performance of different layout and traversal orders. 
We see that \alignedPost traversal performs the best. Indeed, \systemSolver picks the \alignedPost representation, whereas, \systemGreedy chooses the 
\alignedPre representation.

\subsubsection{Righmost}\label{subsec:eval:trees:rightmost}

\begin{figure}[t]
	\includegraphics[width=0.4\linewidth]{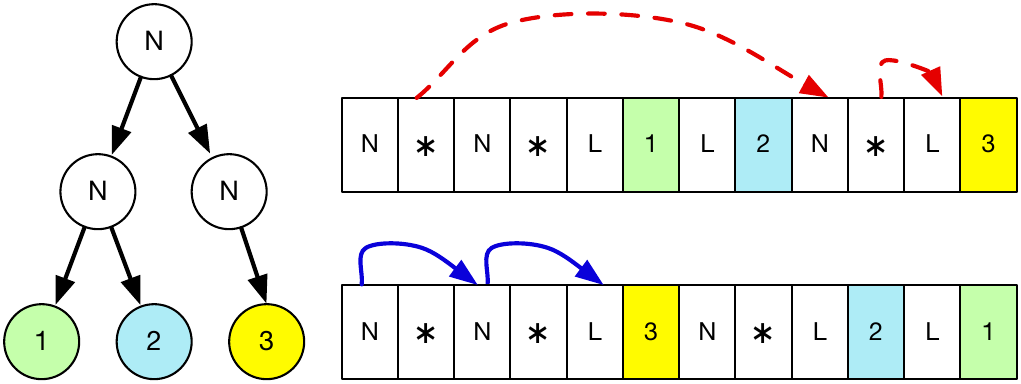}
	\caption{Rightmost: access patterns for two different serializations, left-to-right (top) and right-to-left (bottom)}
	\label{fig:treeShapedData}
\end{figure}

This traversal does recursion on the right child of the tree and returns the \lstinline|Int| value stored in the right-most leaf of the tree.
Figure~\ref{fig:treeShapedData} shows an example of a tree with two different serializations of the tree: left-to-right (top) and right-to-left (bottom).
The right-to-left serialization is more efficient because the constant-step movements (blue arrows) are usually more favorable than variable-step ones (red arrows) on modern hardware.
Both \systemSolver and \systemGreedy pick the right-to-left serialization, and
Table~\ref{table:Right-Most-Tree} shows that this choice performs better in the benchmark.

\begin{table}[]
	\setlength{\tabcolsep}{2pt} \centering
	\ra{1}
	\caption{Rightmost tree runtime statistics in nano seconds (ns).
	Legend: l - left subtree, r - right subtree of the mono tree.}
	\begin{tabular}{@{}p{0.6in}rcrcrcrcrcrcrcr@{}}\toprule 
	 & \multicolumn{3}{|c|}{{\gibbon}} &
	 & \multicolumn{3}{c|}{{\system}}\\ \midrule
		
	 & \multicolumn{1}{c}{{lr}} & 
	 & \multicolumn{1}{c}{{rl}} &
	 & \multicolumn{1}{c}{\small{\systemGreedy}} & 
	 & \multicolumn{1}{c}{\small{\systemSolver}}\\ \midrule
	Mean   
	         &   \textcolor{red}{{383.2}}
			 &&   {318.1} 
			 &&   \textcolor{blue}{{312.2}}
			 &&   {315.4}

			\\
	Median
	        &  \textcolor{red}{{360.0}}
			&&   \textcolor{blue}{{300.0}}
			&&   \textcolor{blue}{{300.0}}
			&&   \textcolor{blue}{{300.0}}
			\\
	ub          
	        &    {404.5}
			&&   {328.9}
			&&   {321.1}
			&&   {327.1}
			\\
	lb           
	        &   {362.4}
			&&  {307.2}
			&&  {303.4}
			&&  {303.8}
			\\ 
	\bottomrule
	\end{tabular}
	\label{table:Right-Most-Tree}
\end{table}

\subsection{Blog Software Case Study}
\label{subsect:blog}

The Blog software case study serves as an example of a realistic benchmark, representing a sample of components 
from a \new{blog management web-service}. The main data structure \new{is a linked-list of blogs where each blog
contains various fields} such as header, id, author information, content, hashtags and date. The fields are a mix
of recursive and non-recursive datatypes. For instance, \lstinline{Content} is a recursive type 
\new{(a \textit{Pandoc} \lstinline|Block| type)}, but \lstinline{Author} is just a single string wrapped in a data
constructor\csk{clarify reasoning packed/non-packed fields.}. Below we show one possible permutation of fields contained in each blog.
\begin{lstlisting}[ ]
data Blogs = Empty | HIADCTB Header Id Author Date Content HashTags Blogs
\end{lstlisting} 
\new{
We evaluate \system{}'s performance using three different traversals over a list of blogs.
Overall, these traversals only use three fields: \lstinline{Content} and \lstinline{Tags}, and the tail of the linked-list, \lstinline{Blogs}.
The fields used by each individual traversal are referred as {\em active fields}
and the rest are referred as {\em passive fields}, and we specify these per-traversal in the following.
In Table~\ref{table:blog-traversals} we report the performance of all six layouts obtained by
permuting the order of these three fields, and two additional layouts (Columns 1 and 2).
The column names are indicative of the order of fields used.
For example, the column \lstinline{hiadctb} reports numbers for a layout where the fields are ordered as:
\lstinline{Header}, \lstinline{Id}, \lstinline{Author}, \lstinline{Date}, \lstinline{Content}, \lstinline{HashTags}, and \lstinline{Blogs} (the tail of the list).
We use \gibbon{} to gather these run times.
The second to last and the last two columns show the run times of these traversals using \system's greedy and solver based optimization.
The layout it picks is tailored to each individual traversal so we specify it per-traversal next.
}

\begin{table}[]
\small
\setlength{\tabcolsep}{2pt} \centering
\ra{1}
\caption{Runtime statistics in seconds (s) for different blog traversals for a subset permutation of different layouts when compiled with \gibbon.
 Legend: h - Header, t - HashTags, b - Blogs, i - TagID, c - Content, a - Author, d - Date.}
\begin{tabular}{@{}p{0.6in}rcrcrcrcrcrcrcrcr@{}}\toprule
 & \multicolumn{13}{|c|}{{\gibbon}} &
 & \multicolumn{3}{c|}{{\system}}\\ \midrule

 & \multicolumn{1}{c}{\small{{hiadctb}}} & 
 & \multicolumn{1}{c}{\small{{ctbhiad}}} &
 & \multicolumn{1}{c}{\small{{tbchiad}}} &
 & \multicolumn{1}{c}{\small{{tcbhiad}}} &
 & \multicolumn{1}{c}{\small{{btchiad}}} &
 & \multicolumn{1}{c}{\small{{bchiadt}}} &
 & \multicolumn{1}{c}{\small{{cbiadht}}} &
 & \multicolumn{1}{c}{\small{\systemGreedy}} &
 & \multicolumn{1}{c}{\small{\systemSolver}}\\ \midrule
Filter\:Blogs \\
Mean   & {\fnumthree{0.23686566388888883}}  
        &&  \fnumthree{0.22362606481481484}
		&&  {\fnumthree{0.05933183453703704}}
		&&  \fnumthree{0.25507014074074075}
		&& {\fnumthree{0.27588248703703705}}
		&&  \textcolor{red}{\fnumthree{0.2878194796296296}}
		&&  \fnumthree{0.20251989074074075} 
		&&   \textcolor{blue}{\fnumthree{0.06728697074074075}} 
		&&  {{{0.071}}}
		\\
Median &  {\fnumthree{0.233412}}  
        &&  \fnumthree{0.2169832}
		&&  {\fnumthree{0.05623404}}
		&&  \fnumthree{0.25230655}
		&&  {\fnumthree{0.27241275}}
		&&  \textcolor{red}{\fnumthree{0.2837016}}
		&&  \fnumthree{0.1973815} 
		&&   \textcolor{blue}{\fnumthree{0.064556175}} 
		&&  \textcolor{blue}{{0.065}}
		\\
ub            &  \fnumthree{0.23839751597495432}
              &&  \fnumthree{0.22587465524096031}
			  &&  \fnumthree{0.06117523036315677}
			  &&  \fnumthree{0.25638477609734245}
			  &&  \fnumthree{0.27777173338789124}
			  &&  \fnumthree{0.2896065944918491}
			  &&  \fnumthree{0.2052628772843204}
			  &&   \fnumthree{0.06880497301300671} 
			  &&  \fnumthree{0.07408468749224115}
			   \\
lb        &  \fnumthree{0.23533381180282334}
              &&  \fnumthree{0.22137747438866937}
			  &&  \fnumthree{0.05748843871091731}
			  &&  \fnumthree{0.25375550538413905}
			  &&  \fnumthree{0.27399324068618286}
			  &&  \fnumthree{0.2860323647674101}
			  &&  \fnumthree{0.1997769041971611}
			  &&   \fnumthree{0.06576896846847478} 
			  &&  \fnumthree{0.06809503121146254}
			   \\
			   \\
Emph\:Content\\
Mean   & \fnumthree{0.6768964925925925}  
        &&  \fnumthree{0.6450119361111112}
		&&  {\fnumthree{1.2521705555555556}}
		&&  \fnumthree{0.6615218694444444}
		&& \fnumthree{1.247634222222222}
		&&  \textcolor{red}{\fnumthree{1.2690613333333334}} 
		&&  {\fnumthree{0.4834597694444444}}
		&&   \textcolor{blue}{\fnumthree{0.481856937037037}}
		&&  {{0.648}}
		\\
Median &  \fnumthree{0.6646339}
        &&  \fnumthree{0.62971775}
		&&  {\fnumthree{1.247218}}
		&&  \fnumthree{0.6474352000000001}
		&&  \fnumthree{1.246326}
		&&  \textcolor{red}{\fnumthree{1.271372}} 
		&&  {\fnumthree{0.46485665}}
		&&  \textcolor{blue}{\fnumthree{0.46374265000000003}}  
		&&  {{0.646}}
		\\
ub        &  \fnumthree{0.684214375971199}
              &&  \fnumthree{0.6534808814829572}
			  &&  \fnumthree{1.2593180020433652}
			  &&  \fnumthree{0.6696055093753548}
			  &&  \fnumthree{1.253413298510475}
			  &&  \fnumthree{1.2765227205108818}
			  &&  \fnumthree{0.4940382047865605}
			  &&   \fnumthree{0.49257226340003385} 
			  &&  \fnumthree{0.6507161519136292}
			   \\
lb         &  \fnumthree{0.669578609213986}
              &&  \fnumthree{0.6365429907392651}
			  &&  \fnumthree{1.245023109067746}
			  &&  \fnumthree{0.6534382295135339}
			  &&  \fnumthree{1.2418551459339693}
			  &&  \fnumthree{1.261599946155785}
			  &&  \fnumthree{0.4728813341023283}
			  &&   \fnumthree{0.4711416106740402} 
			  &&  \fnumthree{0.6451285046520271}
			   \\
			   \\
Tag\:Search \\
Mean   &  \fnumthree{2.032400527777778}  
        &&  \fnumthree{2.017655231481481}
		&&  \fnumthree{2.7684843333333333}
		&&  {\fnumthree{1.7166548333333334}}
		&&  {\fnumthree{2.802108222222222}}
		&&  \textcolor{red}{\fnumthree{2.809880222222222}}
		&&  \fnumthree{1.8571337407407407} 
		&&   \fnumthree{1.8014907037037033} 
		&&  \textcolor{blue}{{1.733}}
		\\
Median &  \fnumthree{1.99322}  
        &&  \fnumthree{1.9764305}
		&&  \fnumthree{2.766269}
		&&  {\fnumthree{1.661759}}
		&&  {\fnumthree{2.801901}}
		&&  \textcolor{red}{\fnumthree{2.809493}}
		&&  \fnumthree{1.804684} 
		&&   \fnumthree{1.7467454999999998} 
		&&  \textcolor{blue}{{1.724}}
		\\
ub         &  \fnumthree{2.056570170215674}
              &&  \fnumthree{2.0423522418758306}
			  &&  \fnumthree{2.7737697271320148}
			  &&  \fnumthree{1.7495511725668056}
			  &&  \fnumthree{2.805736222347774}
			  &&  \fnumthree{2.8118504354494607}
			  &&  \fnumthree{1.8885287561134954}
			  &&   \fnumthree{1.8344839545429246} 
			  &&  \fnumthree{1.7464097507186982}
			   \\
lb        &  \fnumthree{2.008230885339882}
              &&  \fnumthree{1.9929582210871315}
			  &&  \fnumthree{2.763198939534652}
			  &&  \fnumthree{1.6837584940998611}
			  &&  \fnumthree{2.79848022209667}
			  &&  \fnumthree{2.8079100089949836}
			  &&  \fnumthree{1.7684974528644821}
			  &&   \fnumthree{1.719111522008574} 
			  &&  \fnumthree{1.6333495372739286}
			   \\
\bottomrule
\end{tabular}
\label{table:blog-traversals}
\end{table}

\subsubsection{{\bf FilterBlogs}}

\new{
This traversal filters the list of blogs and only retains those which contain the given
hash-tag in their \lstinline{HashTags} field.\csk{should describe what HashTags field is above.}
}
Since \lstinline{Blogs}, \lstinline{Content}, and \lstinline{HashTags} are recursive fields, changing their layout should represent greater differences in performance.
\new{The {\em active fields} for this traveral are \lstinline{HashTags} and \lstinline{Blogs}.}
Theoretically, the performance of this traversal is optimized when the \lstinline{HashTags} field is serialized before \lstinline{Blogs} on account of the first access to \lstinline{HashTags} in the traversal.
\new{This is confirmed in practice with the layout \textbf{\textit{tbchiad}} being the fastest.}
The access to HashTags comes first since its serialized first in memory and once HashTags is fully traversed, the traversal can access blogs in an optimal manner without unnecessary movement in the packed buffer.

Indeed, the layout chosen by \system puts \lstinline{HashTags} first followed by \lstinline{Blogs}.
\new{
  The order of the rest of the fields remains unchanged compared to the source program,
  but this has no effect on performance since they are {\em passive fields}.
}
As shown in Table~\ref{table:blog-traversals}, the performance of the layout chosen by \system is similar to the layout \textbf{\textit{tbchiad}} when compiled with \gibbon.
Both \systemSolver and \systemGreedy pick the layout \textbf{\textit{tbhiadc}} when compiled from the initial layout \textbf{\textit{hiadctb}}.

\subsubsection{{\bf EmphContent}}

This traversal takes a keyword and a list of blogs as input. It then searches the content of each blog for the keyword and emphasizes all its occurrences if present.
\new{If the keyword isn't present, the content remains unchanged.}
The active fields in this traversal are \lstinline{Content} and \lstinline{Blogs}. Therefore, based on the access pattern (\lstinline{Content} accessed before \lstinline{Blogs}), 
the layout with the best performance should place \lstinline{Content} first followed by \lstinline{Blogs}. Indeed, the layout with the best performance when compiled with \gibbon is \textbf{\textit{cbiadht}}. 
The layout that \systemSolver chooses prioritizes the placement of \lstinline{Blogs} before \lstinline{Content}.
It subsequently changes the traversal to recurse on the blogs first and then emphasize content.
The other passive fields are all placed afterwards.
The layout chosen by \systemSolver is \textbf{\textit{bchiadt}}, whereas the layout chosen by \systemGreedy is \textbf{\textit{cbhiadt}} when compiled from the initial layout \textbf{\textit{hiadctb}}.
Table~\ref{table:blog-traversals} shows the performance of all the layouts for this traversal. The performance of \systemGreedy and
\systemSolver differ because data constructors not shown differ in their layout choices.

\subsubsection{{\bf TagSearch}}

This traversal takes a keyword and a list of blogs as input. It looks for the presence of a keyword in the \lstinline{HashTags} field.
If it is present, it emphasizes this keyword in the \lstinline{Content}, and leaves it unchanged otherwise.
The layout that gives the best performance when compiled with \gibbon is \textbf{\textit{tcbhiad}}. This is as expected based on the 
access patterns of the traversal which access \lstinline{HashTags} followed by \lstinline{Content} followed by \lstinline{Blogs}.
\systemSolver chooses the layout \textbf{\textit{tbchiad}} which places \lstinline|HashTags|, followed by \lstinline|Blogs|, 
followed by \lstinline|Content| and changes the traversal to recurse on \lstinline|Blogs| first and later emphasize
\lstinline|Content| in the \lstinline|then| branch. On the other hand, the layout chosen by \systemGreedy is \textbf{\textit{tcbhiad}}
when compiled from the initial layout \textbf{\textit{hiadctb}}.

\subsubsection{\bf Globally optimizing multiple functions}

We use \system to globally optimize the three blog traversals we discussed above such that we pick one layout for all traversals that minimizes the overall runtime.
\vs{TODO: add layout of solver and greedy version.}
Table~\ref{table:global-blog-traversals} shows the runtime for a layout we compiled using \gibbon (\textit{\textbf{hiadctb}}), \systemSolver(\textit{\textbf{tbchiad}}) 
and \systemGreedy(\textit{\textbf{tbchiad}}). We see that \systemGreedy and \systemSolver do a good job in reducing the traversal time globally. 
All the three traversals are run in a pipelined fashion sequentially. Although, \systemSolver does worse with \textit{TagSearch} when run in a pipelined manner, 
it is actually better performing with \systemSolver when run alone as seen in table~\ref{table:blog-traversals}. 
Note that \systemSolver changes more that one data constructor based on the \textit{inlineable} attribute that \systemGreedy does not.
For instance, \systemSolver uses a packed \lstinline|Inline| list in the \lstinline|Content| with the tail serialized before \lstinline|Inline|.
Whereas, \systemGreedy uses a conventional packed list. In a pipelined execution of the traversals, although this helps reduce the runtime in the case 
of the content search traversal, it inadvertently increases the runtime in case of the tag search traversal due to a cache effect
that can benefit from runtime information.

\begin{table}
\setlength{\tabcolsep}{2pt} \centering
\ra{1}
\caption{Runtime in seconds (s) for different blog traversals in a pipeline when we use \system to globally optimize the data layout.
           The input parameters are different from the single function case.}
\begin{tabular}{@{}p{0.6in}rcrcrcrcrcrcrcr@{}}\toprule
 & \multicolumn{1}{c}{\small{\gibbon}} &
 & \multicolumn{1}{c}{\small{\systemGreedy}} &
 & \multicolumn{1}{c}{\small{\systemSolver}}\\ \midrule
Filter\:Blogs \\
Mean  & \textcolor{red}{\fnumthree{1.9269188888888888}}  
        &&  \textcolor{blue}{\fnumthree{0.10196821313131313}}
		&&  {\fnumthree{0.10424667777777777}}
		\\
Median &  \textcolor{red}{\fnumthree{1.92558}}  
        &&  \textcolor{blue}{\fnumthree{0.1011134}}
		&&  {\fnumthree{0.1027659}}
		\\
ub         &  \fnumthree{1.9307990593114182}
              &&  \fnumthree{0.10260698685757125}
			  &&  \fnumthree{0.10512578108159064}
			   \\
lb         &  \fnumthree{1.9230387184663595}
              &&  \fnumthree{0.10132943940505501}
			  &&  \fnumthree{0.1033675744739649}
			   \\
			   \\
Emph\:Content \\
Mean   & \textcolor{red}{\fnumthree{1.5824796565656565}}
        &&   {\fnumthree{1.3882322727272727}}
		&&  \textcolor{blue}{\fnumthree{1.3257011414141415}}
		\\
Median &  \textcolor{red}{\fnumthree{1.578678}}
        && {\fnumthree{1.38418}}
		&& \textcolor{blue}{\fnumthree{1.320468}}
		\\
ub         &  \fnumthree{1.5892854326549597}
              &&  \fnumthree{1.3962014644193474}
			  &&  \fnumthree{1.3360019268094907}
			   \\
lb         &  \fnumthree{1.5756738804763533}
              &&  \fnumthree{1.380263081035198}
			  &&  \fnumthree{1.3154003560187923}
			   \\
			   \\
Tag\:Search \\
Mean   &  {\fnumthree{2.1865495656565654}}  
        &&  \textcolor{blue}{\fnumthree{1.8305267272727275}}
		&&  \textcolor{red}{\fnumthree{2.3562939292929292}}
		\\
Median &  {\fnumthree{2.182541}}
        &&   \textcolor{blue}{\fnumthree{1.826959}}
		&& \textcolor{red}{\fnumthree{2.351989}}
		\\
ub        &  \fnumthree{2.193582984981658}
              &&  \fnumthree{1.8383555255199808}
			  &&  \fnumthree{2.3646218535316708}
			   \\
lb         &  \fnumthree{2.1795161463314727}
              &&  \fnumthree{1.8226979290254741}
			  &&  \fnumthree{2.3479660050541877}
			   \\
\bottomrule
\end{tabular}
\label{table:global-blog-traversals}
\end{table}

\subsubsection{{\bf Comparison of \system against \mlton}}
\label{subsec:mlton}

We compare \system{}'s performance to \mlton{}, which compiles programs 
written in Standard ML, a strict language, to executables that are small
with fast runtime performance.  

\begin{figure}
	\centering
	\begin{subfigure}[]{0.3\textwidth}
		\centering
		\includegraphics[width=\textwidth]{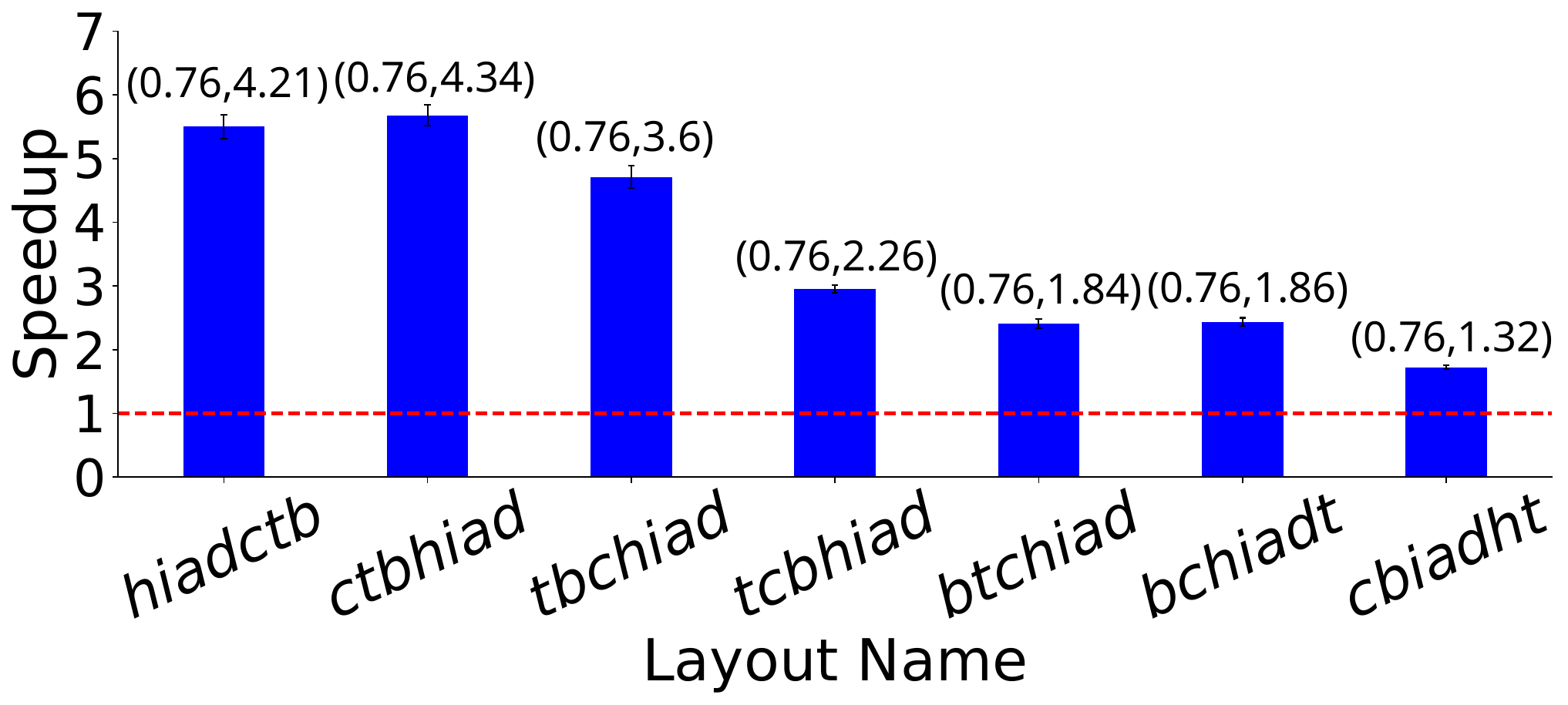}
		\caption{Filter Blogs Traversal.}
		\label{table:speedup-mlton-filterblogs}
	\end{subfigure}
	\hfill
	\begin{subfigure}[]{0.3\textwidth}
		\centering
		\includegraphics[width=\textwidth]{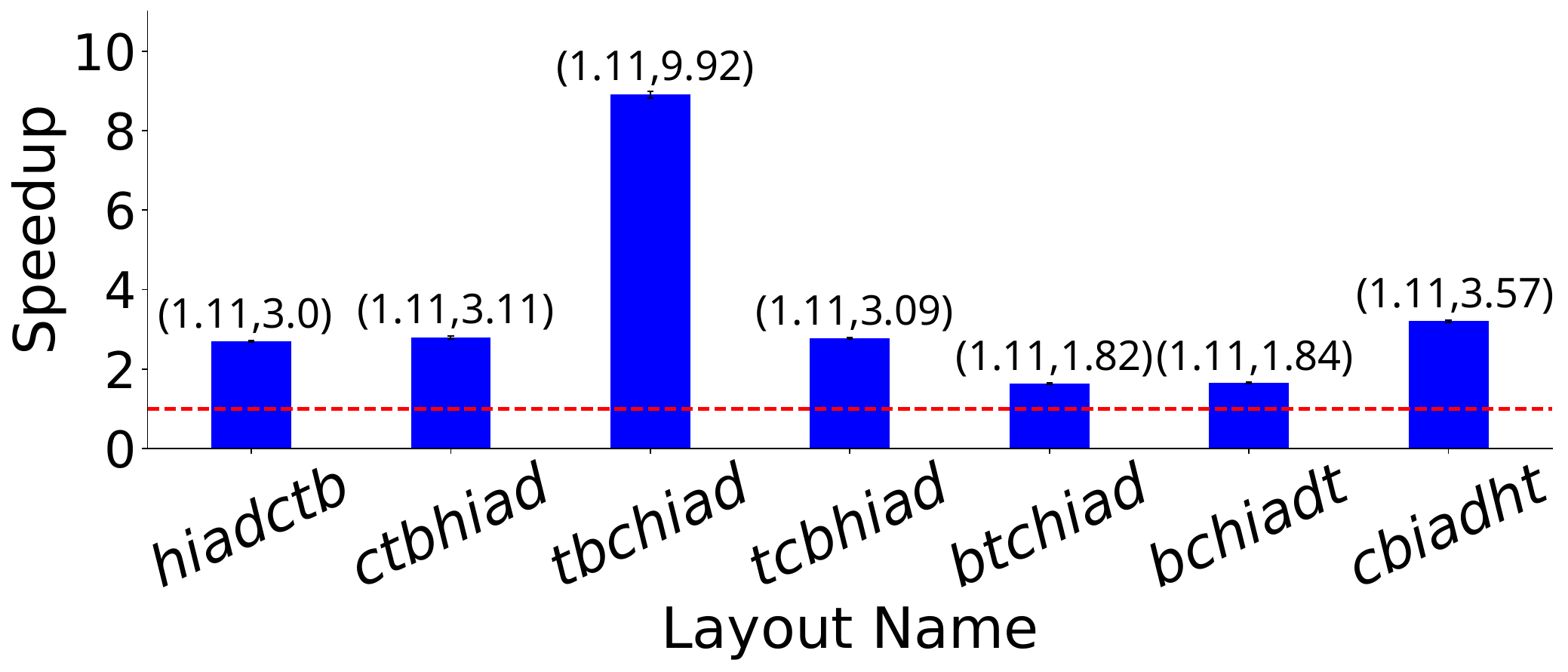}
		\caption{Content Search Traversal.}
		\label{table:speedup-mlton-contentsearch}
	\end{subfigure}
	\hfill
	\begin{subfigure}[]{0.3\textwidth}
		\centering
		\includegraphics[width=\textwidth]{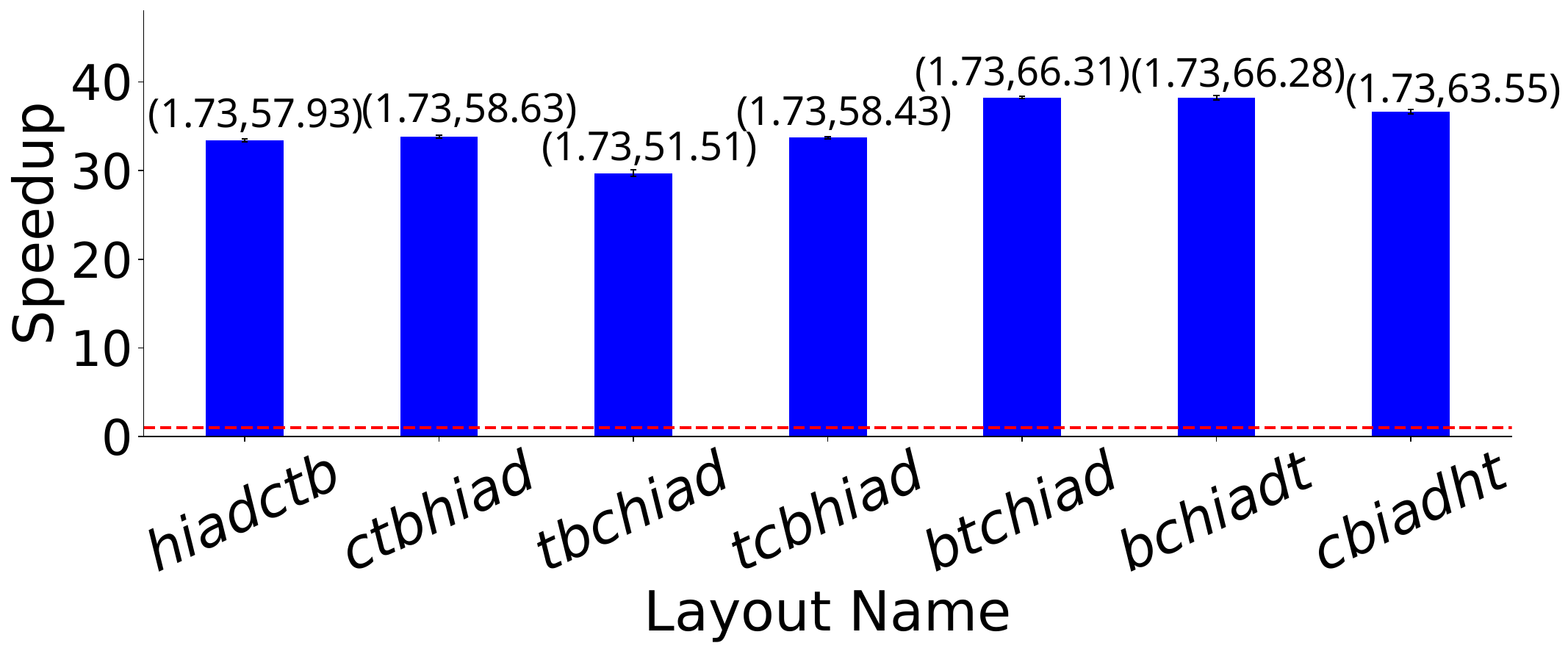}
		\caption{Tag Search Traversal.}
		\label{table:speedup-mlton-tagsearch}
	\end{subfigure}
	   \caption{Performance comparison of \systemSolver with \mlton. 
	   The tuple on top of each bar shows the median runtime in seconds of \systemSolver followed by the corresponding layout when compiled with \mlton.}
	   \label{fig:speedupghc}
\end{figure}

\Figref{fig:speedupghc} shows the the speedup of \system over \mlton.
As shown, the performance of \system is better than \mlton by significant margins for all the layouts and traversals.
Since ADTs in \mlton are \textit{boxed} -- even though native integers or native arrays are \textit{unboxed} -- such a behavior is expected 
because it adds more instructions (pointer de-referencing) and results in worse spatial locality.

\subsection{Cache behavior}\label{subsec:cache}

\begin{table}
    \small
	\setlength{\tabcolsep}{2pt} \centering
	\ra{1}
	\caption{PAPI performance counter statistics (average of 99 runs) for different blog traversals.}
	\begin{tabular}{@{}p{0.6in}rcrcrcrcrcrcrcrcr@{}}\toprule
	  & \multicolumn{13}{|c|}{{\gibbon}} &
      & \multicolumn{3}{c|}{{\system}}\\ \midrule

	 & \multicolumn{1}{c}{\small{{hiadctb}}} & 
	 & \multicolumn{1}{c}{\small{{ctbhiad}}} &
	 & \multicolumn{1}{c}{\small{{tbchiad}}} &
	 & \multicolumn{1}{c}{\small{{tcbhiad}}} &
	 & \multicolumn{1}{c}{\small{{btchiad}}} &
	 & \multicolumn{1}{c}{\small{{bchiadt}}} &
	 & \multicolumn{1}{c}{\small{{cbiadht}}} &
	 & \multicolumn{1}{c}{\small{{\systemGreedy}}} &
	 & \multicolumn{1}{c}{\small{\systemSolver}}\\ \midrule
	Filter\:Blogs\\
	Ins    & {\snum{583007860.8888888}} &&  \snum{582007417.2222222} && {\snum{583006557.5555556}}   && {\snum{581007422.0}} && \textcolor{blue}{\snum{579006542.1111112}} && \snum{578007431.3333334} &&   \textcolor{red}{\snum{586007412.4444444}} && \snum{583006551.4444444} &&   \snum{583006570.4444444}    \\
	Cycles    & {\snum{989222808.8888888}} &&  \snum{959711380.1111112} && \textcolor{blue}{\snum{285081123.8888889}}   && {\snum{1052921958.2222222}} && {\snum{1200439241.3333333}} && \textcolor{red}{\snum{1248215771.1111112}} &&   {\snum{877792663.1111112}} && \snum{279146331.4444444} &&   \snum{289017037.7777778}    \\
	L2\:DCM    & {\snum{11185155.333333334}} &&  \snum{11510078.222222222} && {\snum{938489.3333333334}}   && \textcolor{red}{\snum{13337049.333333334}} && {\snum{12928602.222222222}} && \snum{13065840.0} &&   {\snum{7789525.777777778}} && \snum{889989.7777777778} &&   \textcolor{blue}{\snum{874607.7777777778}}    \\
	\\
	Content\:Search\\
	Ins    & {\snum{5845022229.888889}} &&  \snum{5834021360.444445} && {\snum{6776362325.666667}}   && {\snum{5841021767.888889}} && {\snum{6779241706.444445}} && \textcolor{red}{\snum{6779244615.888889}} &&   \textcolor{blue}{\snum{5837021292.777778}} && \snum{5837021295.333333} &&   \snum{5837021366.222222}    \\
	Cycles   & {\snum{2892647565.888889}} &&  \snum{2738456405.4444447} && {\snum{4272337912.0}}   && {\snum{2842427055.6666665}} && \textcolor{red}{\snum{4340585270.333333}} && \snum{4291414051.5555553} &&   \textcolor{blue}{\snum{2056621185.3333333}} && \snum{2061807292.6666667} &&   \snum{2812417391.5555553}    \\
	L2\:DCM    & {\snum{12800140.0}} &&  \snum{10780195.333333334} && {\snum{20295265.222222224}}   && {\snum{13283107.333333334}} && \textcolor{red}{\snum{20549853.111111112}} && \snum{20994551.666666668} &&   \textcolor{blue}{\snum{7727582.888888889}} && \snum{7663269.0} &&   \snum{10702561.555555556}    \\
	\\
	Tag\:Search \\
	Ins    & {\snum{22490860539.77778}} &&  \textcolor{blue}{\snum{22487260185.555557}} && \textcolor{red}{\snum{23043358250.333332}}   && {\snum{22487260291.333332}} && {\snum{23043009879.0}} && \snum{23041811630.22222} &&   {\snum{22488060134.77778}} && \snum{22487659963.444443} &&   \snum{22487659965.0}    \\
	Cycles    & {\snum{8587261902.555555}} &&  \snum{8593729939.88889} && {\snum{9609104561.222221}}   && \textcolor{blue}{\snum{7288093434.666667}} && {\snum{9607367711.666666}} && \textcolor{red}{\snum{9748577539.11111}} &&   {\snum{7883183997.666667}} && \snum{7609921682.888889} &&   \snum{7568236436.222222}    \\
	L2\:DCM    & {\snum{20227011.777777776}} &&  \snum{20621886.333333332} &&  \textcolor{red}{\snum{39956792.44444445}}   && \textcolor{blue}{\snum{10612695.222222222}} && {\snum{28602975.777777776}} && \snum{26479819.222222224} &&  {\snum{16392591.777777778}} && \snum{12885043.444444444} &&   \snum{12753814.333333334}    \\
	\bottomrule
	\end{tabular}
	\label{table:filterblogs-papi}
	\end{table}

The results from earlier sections demonstrate that \system's layout choices improve run-time performance. This section investigates {\em why} performance improves. The basic premise of \system's approach to layout optimization is to concentrate on minimizing how often a traversal needs to backtrack or skip ahead while processing a buffer. By minimizing this jumping around, we expect to see improvements from two possible sources. First, we expect to see an improvement in instruction counts, as an optimized layout should do less pointer chasing. Second, we expect to see an improvement in L2 and L3 cache utilization: both fewer misses (due to improved spatial locality and prefetching) and fewer accesses (due to improved locality in higher level caches).

Table~\ref{table:filterblogs-papi} shows that our main hypothesis is borne out and the optimal layout has fewer
L2 data cache misses\footnote{Since PAPI, the processor counters framework, does not completely support latest AMD processors yet, we were unable to obtain L3 cache misses, only the L2 DCM counter was available.}:
a better layout promotes better locality. Interestingly, we do not observe a similar effect for instruction count. While different layouts differ in instruction counts, the difference is slight. We suspect this light effect of a better layout may be due to \gibbon's current implementation, which often dereferences pointers even if a direct access in the buffer would suffice.

\subsection{Tradeoffs between \system's solver and greedy optimization}

\begin{figure}
	\centering
	\begin{subfigure}[b]{0.3\textwidth}
		\centering
		\includegraphics[width=\textwidth]{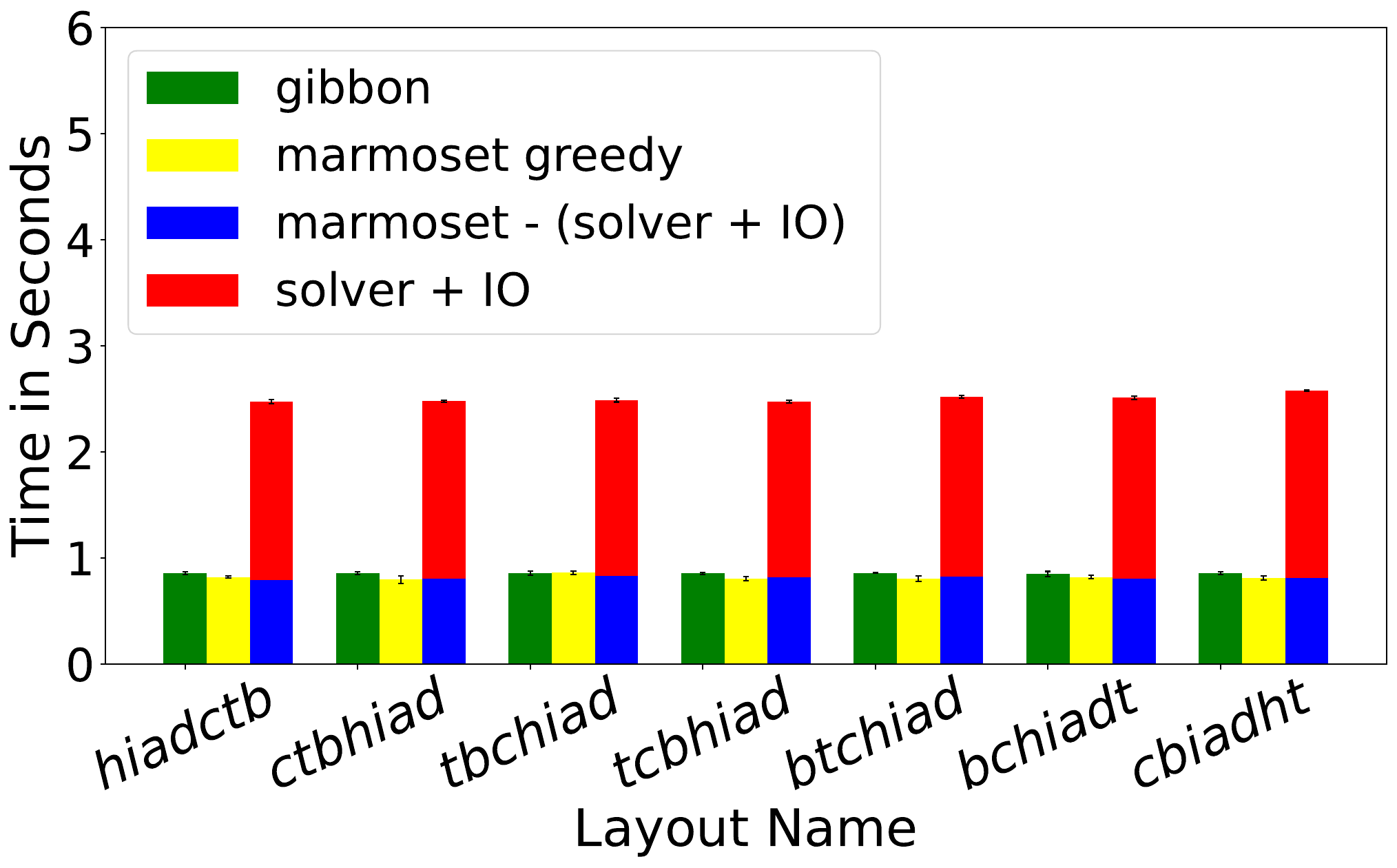}
		\caption{Compile times for the Filter blogs traversal.}
		\label{table:compiletimes-filterblogs}
	\end{subfigure}
	\hfill
	\begin{subfigure}[b]{0.3\textwidth}
		\centering
		\includegraphics[width=\textwidth]{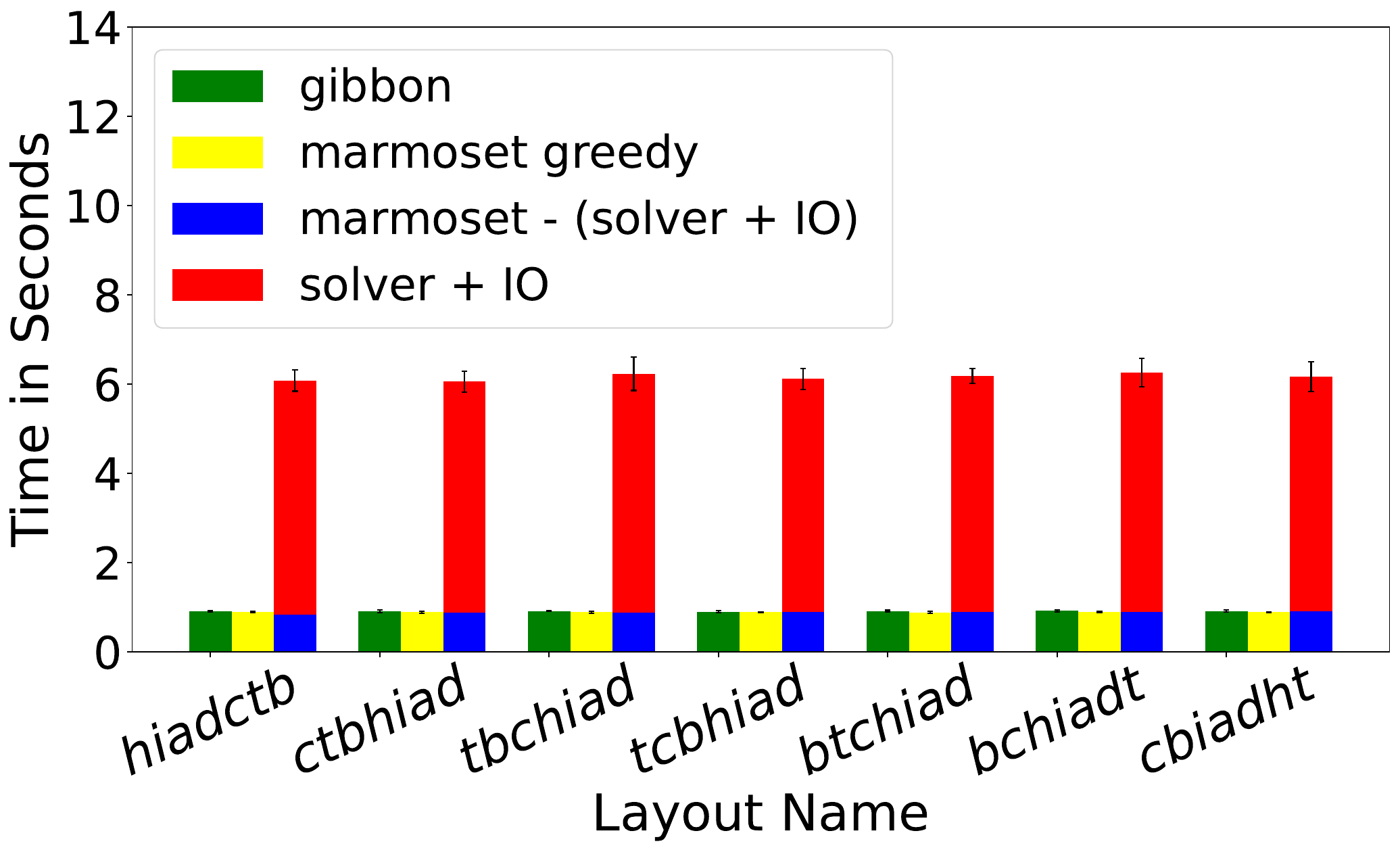}
		\caption{Compile times for the Content Search traversal.}
		\label{table:compiletimes-contentsearch}
	\end{subfigure}
	\hfill
	\begin{subfigure}[b]{0.3\textwidth}
		\centering
		\includegraphics[width=\textwidth]{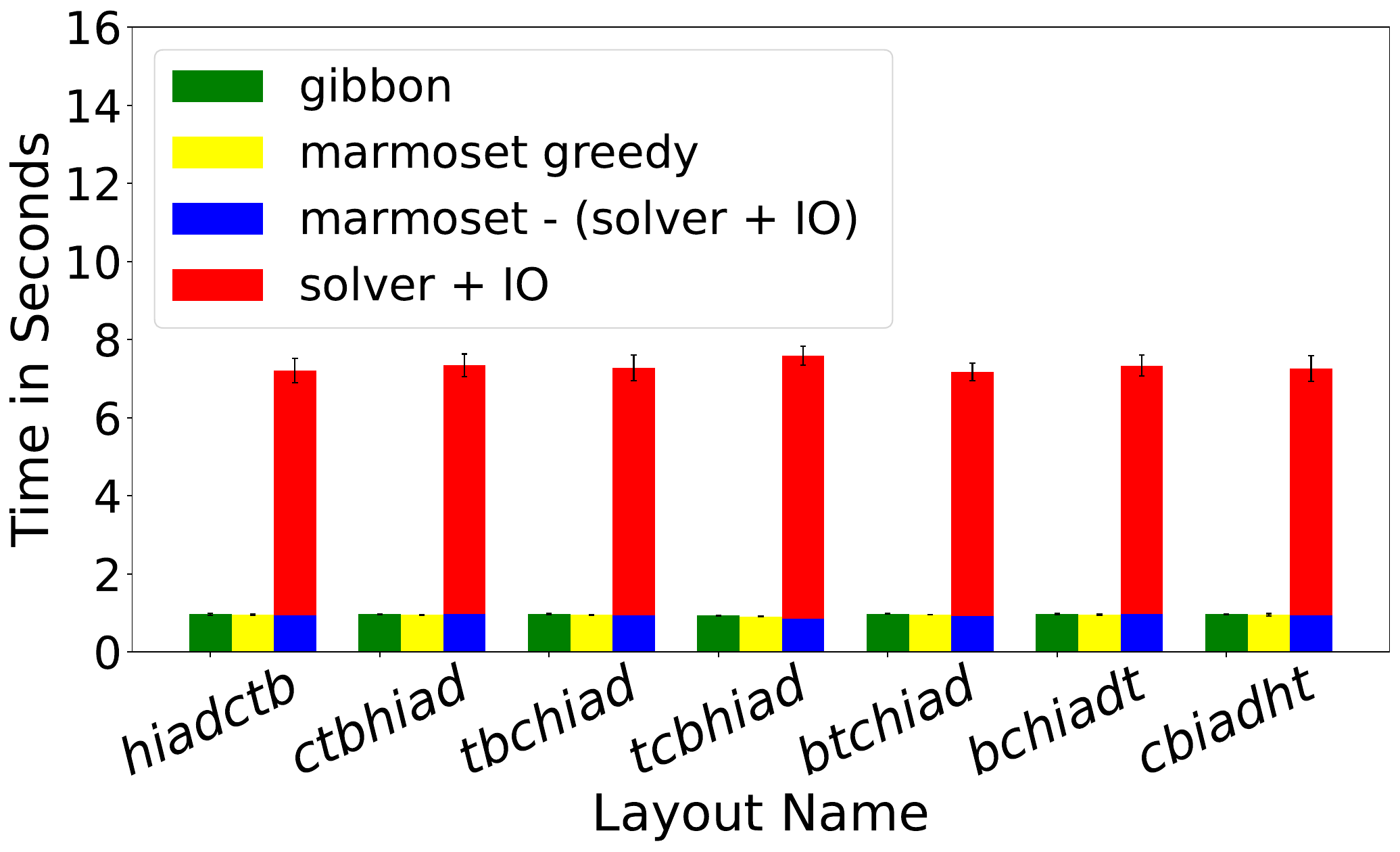}
		\caption{Compile times for the Tag Search traversal.}
		\label{table:compiletimes-tagsearch}
	\end{subfigure}
	   \caption{Average compile times (99 runs) in seconds for different layouts and traversal combinations
	   			 when compiled with \gibbon and when optimized for layout by \systemGreedy and \systemSolver.
				 The bar in \textit{Blue} is the compile time in seconds for \systemSolver after subtracting the
				 time for solver IO and the bar in \textit{Red} is the time taken to do the IO call to the
				 DOCPLEX solver. The total compile time for \systemSolver includes the time for doing the CFG
				 analysis, generating the access graph, the solver time, reordering the datatype and repairing
				 the code, The majority of which is added by the solver IO.
				 }
	   \label{table:compiletimes}
\end{figure}

To understand the difference between the layout chosen by \systemSolver and \systemGreedy we now take a closer look at the tag search traversal shown in Table~\ref{table:blog-traversals}.
Here, both versions choose two different layouts with different performance implications. \systemSolver chooses the layout \textbf{\textit{tbchiad}} whereas \systemGreedy chooses 
the layout \textbf{\textit{tcbhiad}}. The mechanics of why can be explained using our running example (Figure~\ref{fig:blog-traversal}) which is essentially a simplified version of the 
traversal shown in the evaluation. Figure~\ref{graph:running-access} shows the access graph for this traversal. \systemSolver generates constraints 
outlined in Section~\ref{design:genconstraints} that lead to the layout \textbf{\textit{tbchiad}}. On the other hand, \systemGreedy starts at the root node of the graph and greedily chooses the next child to traverse. 
The order in which nodes of the graph are visited fixes the order of fields in the data constructor. In this case, the root node is \lstinline|HashTags| which makes it the first field in the greedy 
layout, next, the greedy heuristic picks the \lstinline|Content| field making it the second field and finally followed by the \lstinline|BlogList| field.

We show the compile times in \figref{table:compiletimes} for different layout and traversal combinations when compiled with \gibbon, \systemGreedy and \systemSolver respectively as 
a measure of relative costs. The compile times for \systemSolver include the time to generate the control flow graph, the field access graph, the solver time and the time to re-order the datatype in the code.
The solver times are in the order of the number of fields in a data constructor and not the program size. Hence, the solver adds relatively low overhead.
Since the compiler does an IO call to the python solver, there is room for improvement in the future to lower these times. 
For instance, we could directly perform \textit{FFI} calls to the CPLEX solver by lowering the constraints to C code. 
This would be faster and safer than the current implementation. In addition, during the global optimization, we call the solver on each data constructor as of the moment, 
we could further optimize this by sending constraints for all data constructors at once and doing just one solver call. 

Although the cost of \system's solver based optimization is higher that the greedy approach, it is a complementary approach 
which may help the user find a better layout at the cost of compile time. On the other hand, if the user wishes to optimize 
for the compile time, they should use the greedy heuristic.

\subsection{Discussion: Scale of Evaluation}

\system's approach for finding the best layout for densely presented
data is language agnostic, but the evaluation has to be language specific.
Hence, we implemented the approach inside a most-developed (to our
knowledge) compiler supporting dense representations of recursive datatypes, the
Gibbon compiler. Our evaluation is heavily influenced by this.

At the time of writing, the scale of evaluation is limited by a number of
Gibbon-related restrictions.
Gibbon is meant as a tree traversal accelerator~\cite{gibbon} and
its original suite of benchmarks served as a basis and inspiration for
evaluation of \system.
``Big'' end-to-end projects (e.g. compilers, web servers, etc.)
have not been implemented in Gibbon and, therefore, are out of reach for us.
If someone attempted to implement such a project using Gibbon,
they would have to extend the compiler to support many realistic features:
modules, FFI, general I/O, networking.
Alternatively, one could integrate Gibbon into an existing realistic compiler
as an optimization pass or a plugin.
For instance, the Gibbon repository has some preliminary work for integrating as a
GHC plugin\footnote{\url{https://github.com/iu-parfunc/gibbon/tree/24c41c012a9c33bff160e54865e83a5d0d7867dd/gibbon-ghc-integration}
}, but it is far from completion.
In any case, the corresponding effort is simply too big.
Overall, the current \system evaluation shows that our approach is viable.
\label{subsec:heuristic-tradeoffs}

 \section{Discussion}
\label{sec:discussion}

\ap{make sure to discuss possible improvements that used to be mentioned in Sec
  1,2: pointer elimination, user annotations}
\vs{added improvements mentioned in sec 1,2. Fixed part about annotations.}  
Throughout the case studies, we see that \system
can optimize the layout of datatypes and provide performance gains. We
believe that these gains can be further improved by exploring ideas briefly discussed
in this section.

\system allows the user to provide optional manual constraints on the layout
as either relative or absolute constraints through \textit{ANN} annotation pragmas.
The relative constraint allows the user to specify if a field A comes immediately after
field B. The absolute constraint specifies an exact index in the layout for a field. 
Such pragmas may be beneficial if the user requires a specific configuration over data
types or has information about performance bottlenecks.

Although the performance optimization is currently statically driven, there are many avenues for future
improvement. For instance, we can get better edges weights for the access graphs
using dynamic profiling techniques. The profiling can be quite detailed, for
instance, which branch in a function is more likely, which function takes the
most time overall in a global setting (the optimization would bias the layout
towards that function), how does a particular global layout affect the
performance in case of a pipeline of functions.

We could also look at a scenario where we optimize each function locally and 
use ``shim'' functions that copy one layout to another (the one required by the
next function in the pipeline). Although the cost of copying may be high, it warrants
further investigation. Areas of improvement purely on the implementation side include 
optimizing whether \system dereferences a pointer to get to a field or uses the end-witness
information as mentioned in section~\ref{sec:overview}. Lesser pointer dereferencing 
can lower instruction counts and impact performance positively. We would also like to 
optimize the solver times as mentioned in~\ref{subsec:heuristic-tradeoffs}.

We envision that the \textit{structure of arrays} effect that we discovered may help 
with optimizations such as vectorization, where the performance can benefit significantly
if the same datatype is close together in memory. Regardless, through the case studies, we see 
that \system shows promise in optimizing the layout of datatypes and may open up 
the optimization space for other complex optimizations such as vectorization.
 \section{Related Work}

\ap{oopsla reviewers asked to include Hawkins et al PLDI '11
https://dl.acm.org/doi/abs/10.1145/1993498.1993504}

\subsection{Cache-conscious data}

\citet{ccgarbage} base on an object-oriented language with a
generational garbage collector, which they extend with a heuristic for
copying objects to the TO space.
Their heuristic uses a special-purpose graph data structure, the \emph{object affinity
graph}, to identify when groups of objects are accessed by the program
close together in time.
When a given group of objects have high affinity in the object
affinity graph, the collector is more likely to place them close
together in the TO space.
As such, a goal of their work and ours is to achieve higher data-access locality
by carefully grouping together objects in the heap.
However, a key difference from our work is that their approach bases its placement decisions
on an object-affinity graph that is generated from profiling data, which is
typically collected online by some compiler-inserted instrumentation.
The placement decisions made by our approach are based on data collected by static analysis of the program.
Such an approach has the advantage of not depending on the output of dynamic profiling, and therefore
avoids the implementation challenges of dynamic profiling.
A disadvantage of not using dynamic profiling is that the approach
cannot adapt to changing access patterns that are highly input
specific.
We leave open for future work the possibility of getting the best of both approaches.

\citet{CSD} introduce the idea of hot/cold splitting of a data structure, where elements are categorized as being ``hot'' if accessed frequently and ``cold'' if accessed inferquently.
This information is obtained by profiling the program. Cold fields are placed into a new object via an indirection and hot fields remain unchanged. In their approach, at runtime, there is a cache-concious garbage collector~\cite{ccgarbage} that co-locates the modified object instances.
This paper also suggests placing fields with high temporal affinity within the same cache block. For this they recommend  \texttt{bbcache}, a field recommender for a data structure. \texttt{bbcache} forms a field affinity graph which combines static information about the source location of structure field accesses with dynamic information about the temporal ordering of accesses and their access frequency.

\citet{CSL} propose two techniques to solve the problem of poor reference locality.
\begin{itemize}[leftmargin=*]
\item \textbf{ccmorph:} This works on tree-like data structures, and it relies on the programmer making a calculated guess about the safety of the operation on the tree-like data structure. It performs two major optimizations: clustering and coloring. Clustering take a the tree like data structure and attempts to pack likely to be accessed elements in the structure within the same cache block. There are various ways to pack a subtree, including clustering $k$ nodes in a subtree together, depth first clustering, etc.
Coloring attempts to map simultaneously accessed data elements to non-conflicting regions of the cache.
\item \textbf{ccmalloc:} This is a memory allocator similar to \texttt{malloc} which takes an additional parameter that points to an existing data structure element which is likely accessed simultaneously. This requires programmer knowledge and effort in recognizing and then modifying the code with such a data element. \texttt{ccmalloc} tries to allocate the new data element as close to the existing data as possible, with the initial attempt being to allocate in the same cache block. It tries to put likely accessed elements on the same page in an attempt to improve TLB performance.
\end{itemize}

\citet{Shapes, Shapes2} suggest
that the layout of a data structure should be defined once at the point of initialization, and all further code that interacts with the structure should be ``layout agnostic''. Ideally, this means that performance improvements involving layout changes can be made without requiring changes to program logic. To achieve this, classes are extended to support different laouts, and types carry layout information---code that operates on objects may be polymorphic over the layout details.

\subsection{Data layout description and binary formats}

\citet{Dargent} propose a data layout description framework \emph{Dargent}, which allows programmers to specify the data layout of an ADT. It is built on top of the Cogent language~\cite{o2016cogent}, which is a first order polymorphic functional programming language. Dargent targets C code and provides proofs of formal correctness of the compiled C code with respect to the layout descriptions. Rather than having a compiler attempting to determine an efficient layout, their focus is on allowing the programmer to specify a particular layout they want and have confidence in the resulting C code.

Significant prior work went into generation of verified efficient code for interacting with binary data formats (parsing and validating). For example, EverParse~\cite{EverParse} is a framework for generating verified parsers and formatters for binary data formats, and it has been used to formally verify zero-copy parsers for authenticated message formats. With Narcissus~\cite{Narcissus}, encoders and decoders for binary formats could be verified and extracted, allowing researchers to certify the packet processing for a full internet protocol stack. Other work~\cite{GenericPacketDescriptions} has also explored the automatic generation of verified parsers and pretty printers given a specification of a binary data format, as well as the formal verification of a compiler for a subset of the Procotol Buffer serialization format~\cite{VPBdeleware}.

\citet{DataScript} demonstrates how a domain-specific language for describing binary data formats could be useful for generating validators and for easier scripting and manipulation of the data from a high-level language like Java. \cite{PacketTypes} introduce \emph{Packet Types} for programming with network protocol messages and provide language-level support for features commonly found in protocol formats like variable-sized and optional fields.

\citet{Hawkins} introduce \relc, a framework for synthesizing low-level C\texttt{++} code from a high-level
relational representation of the code. The user describes and writes code that represents data at a high level as relations. Using a decomposition
of the data that outlines memory representation, \relc synthesizes correct and efficient low-level code.

\citet{BitStealingMadeLegal} introduce the \ribbit DSL, which allows programmers to describe the layout of ADTs that are monomorphic and immutable.
\ribbit provides a dual view on ADTs that allows both a high-level description of the ADT that the client code follows and a user-defined memory representation of the ADT
for a fine-grained encoding of the layout. Precise control over memory layout allows \ribbit authors to develop optimization algorithms over the ADTs, such as struct packing, bit stealing, pointer tagging, unboxing, etc.
Although this approach enables improvements to layout of ADTs, it is different from \system's: \ribbit focuses on \emph{manually} defining low-level memory representation of the ADT whereas \system \emph{automatically} optimizes the high-level layout (ordering of fields in the definition ADT) relying on \gibbon for efficient packing of the fields.
While \ribbit invites the programmer to encode their best guess about the optimal layout, \system comes up with such layout by analysing access patterns in the source code.

\subsection{Memory layouts}

Early work on specifications of memory layouts was explored in various studies of PADS, a language for describing ad hoc data-file formats~\cite{PADS, Pads2, Pads3}.

\citet{poolallocation} introduce a technique for improving the memory layout of the heap of a given C program.
Their approach is to use the results of a custom static analysis to enable \emph{pool allocation} of heap objects.
Such automatic pool allocation bears some resemblance to our approach, where we use region-based allocation in tandem with region inference, and thanks to static analysis can group fields of a given struct into the same pool, thereby
improving locality in certain circumstances.

Floorplan~\cite{Floorplan} is a declarative language for specifying high-level memory layouts, implemented as a compiler which generates Rust code. The language has forms for specifying sizes, alignments, and other features of chunks of memory in the heap, with the idea that any correct state of the heap can be derived from the Floorplan specification. It was successfully used to eliminate 55 out of 63 unsafe lines of code (all of the unsafe code relating to memory safety) of the immix garbage collector.

 \section{Conclusions}

\ap{probably needs rewriting for ECOOP resubmission}
This paper introduced \system, which builds on Gibbon to generate efficient orders for Algebraic Datatypes. We observed that a fairly straightforward control flow and data flow analysis allows \system to identify opportunities to place fields of a data constructor near each other in memory to promote efficient consecutive access to those fields. Because a given function might use many fields in many different ways, \system adopts an approach of formulating the data layout problem as an ILP, with a cost model that assigns an abstract cost to a chosen layout. This ILP can be further constrained with user-provided layout guidelines. Armed with problem formulation, an off-the-shelf ILP solver allows \system to generate minimal-(abstract)-cost layouts for algebraic datatypes. \system then uses these layouts to synthesize a new ADT and prior work's Gibbon compiler toolchain to lower the code into an efficient program that operates over packed datatypes with minimal pointer chasing.

We found, across a number of benchmarks, that \system is able to effectively and consistently find the optimal data layout for a given combination of traversal function and ADT. We showed, experimentally, that \system-optimized layouts significantly outperform not only Gibbon's default layouts, but also GHC's standard compiler.

\begin{acks}
\end{acks}

\bibliographystyle{ACM-Reference-Format}

\appendix

\bibliography{refs}

\end{document}